\documentclass[graybox]{svmult}

\usepackage{mathptmx}       
\usepackage{helvet}         
\usepackage{courier}        
\usepackage{type1cm}        
\usepackage{makeidx}         
\usepackage{graphicx}        
\usepackage{multicol}        
\usepackage[bottom]{footmisc}

\usepackage{amssymb,amsmath}

\makeindex             

\begin{document}

\title*{Cosmology with gravitational lensing}
\author{Alan Heavens}
\institute{Alan Heavens \at  Institute for Astronomy, University of Edinburgh, Blackford Hill, Edinburgh EH9 3HJ, U.K., \email{afh@roe.ac.uk}}
%
%
\maketitle

\abstract*{In these lectures I give an overview of gravitational lensing, concentrating on theoretical aspects, including derivations of some of the important results.  Topics covered include the determination of surface mass densities of intervening lenses, as well as the statistical analysis of distortions of galaxy images by general inhomogeneities (cosmic shear), both in 2D projection on the sky, and in 3D where source distance information is available.  3D mass reconstruction and the shear ratio test are also considered, and the sensitivity of observables to Dark Energy is used to show how its equation of state may be determined using weak lensing.  Finally, the article considers the prospect of testing Einstein's General Relativity with weak lensing, exploiting the differences in growth rates of perturbations in different models.}

\abstract{In these lectures I give an overview of gravitational lensing, concentrating on theoretical aspects, including derivations of some of the important results.  Topics covered include the determination of surface mass densities of intervening lenses, as well as the statistical analysis of distortions of galaxy images by general inhomogeneities (cosmic shear), both in 2D projection on the sky, and in 3D where source distance information is available.  3D mass reconstruction and the shear ratio test are also considered, and the sensitivity of observables to Dark Energy is used to show how its equation of state may be determined using weak lensing.  Finally, the article considers the prospect of testing Einstein's General Relativity with weak lensing, exploiting the differences in growth rates of perturbations in different models.}

\font\bfmath=cmmib10
\def\balpha{\hbox{\bfmath\char'013}}
\def\bbeta{\hbox{\bfmath\char'014}}
\def\vth{\hbox{\bfmath\char'022}}    
\newcommand{\bell}{\mbox{\boldmath{$\ell$}}}
\newcommand{\edth}{\,\tilde\partial\,}
\newcommand{\Aff}[3]{\Gamma^{#1}_{{\phantom{#1}{#2}{#3}}}}

\section{Introduction}
 
Gravitational lensing has emerged from being a curiosity of Einstein's General Relativity to a powerful cosmological tool.  The reasons are partly theoretical, partly technological.  The traditional tool of observational cosmology, the galaxy redshift survey, has become so large that statistical errors are very small, and questions about the fundamental limitations of this technique were raised.  Fundamentally, studies of galaxy surveys will be limited by uncertain knowledge of galaxy formation and evolution - we will probably never know with high accuracy where galaxies should exist, even statistically, in a density field.  Since it is the density field which is most directly predictable from fundamental theories (with some caveats), gravitational lensing is attractive as it is a direct probe.  Furthermore, statistical analysis shows that large weak lensing surveys covering large fractions of the sky to a median redshift of around unity are in principle extremely powerful, and can lead to error bars on cosmological parameters which are very small.  In particular, the subtle effects of Dark Energy and even modifications to Einstein's General Relativity are potentially detectable with the sort of surveys which are being planned for the next decade.  The limitations of lensing are likely to be in systematic errors arising from the ability to measure accurately shapes of galaxy images, physical effects aligning galaxies themselves, uncertain distances of sources, and uncertainties in the theoretical distribution of matter on small scales where baryon physics becomes important.   Specially-designed instrumentation and telescopes with excellent image quality help the situation a lot, and understanding of the physical systematics is now quite good. These notes will not be concerned very much with the practical issues, but more with the theoretical aspects of how lensing works, how it can be used to determine surface densities of clusters of galaxies (testing the Dark Matter content), how it can be used statistically on large scales (testing the Dark Energy properties) and how the combination of geometrical measurements and the growth rate of perturbations can probe the gravity law (testing so-called Dark Gravity).  It is not a comprehensive review, and in particular does not cover the observational developments, which have seen the typical size of lensing surveys increase from 1 (past) $\rightarrow$ 100 (now) $\rightarrow  10^4$ (near future) square degrees.  For more comprehensive recent reviews, see e.g. \cite{Munshi08, HJ08} or the excellent SAAS-FEE lecture notes \cite{SAASFEE}.  The structure of these notes is that section 2 covers basic lensing results, section 3 (Dark) Matter mapping, section 4 lensing on a cosmological scale, section 5 3D lensing  and Dark Energy, and section 6 Dark Gravity. 
 
\section{Basics of lensing}

We begin with some basic results on light bending, and derive results for the bending of light by an intervening lens with an arbitrary surface density.

\subsection{The bend angle}

The basic mechanism for gravitational lensing is that a point mass
$M$ will deflect a light beam through an angle
\begin{equation}
\tilde\alpha(r) = {4GM\over r c^2}\qquad {\rm Point\ Mass}\label{alpha}
\end{equation}
where $G$ is Newton's gravitation constant, $c$ is the speed of
light, and $r$ is the distance of closest approach of the ray to the
mass.  This deflection angle is calculated using Einstein's General
Theory of Relativity (GR), but is simply a factor two larger than
Newtonian theory would predict if one treats the photon as a massive
particle.  The light path is curved, but if we are looking at a source which is far
beyond the lens, the light path can be approximated by two straight
lines, with a bend angle $\tilde\alpha$ between them.  This is the {\em thin-lens}
approximation.

The bend angle for a point mass equation (\ref{alpha}) can be used
to calculate the bend angle for an arbitrary mass distribution. For
example, the bend angle for a thin
lens whose mass distribution projected on the sky is
circularly-symmetric depends on the projected mass enclosed
within a circle centred on the middle of the lens, and extending out
to the projected radius $R$ of the ray:
\begin{equation}
\tilde\alpha(R) ={4GM(<R)\over R c^2}\qquad {\rm Circularly-symmetric\
mass}\label{circ}
\end{equation}
where $M(<R)$ is the mass enclosed within a projected radius $R$.

\subsection{The lens equation}

\begin{figure}
\sidecaption
\vspace{4cm}
\scalebox{0.4}{\includegraphics[angle=0]{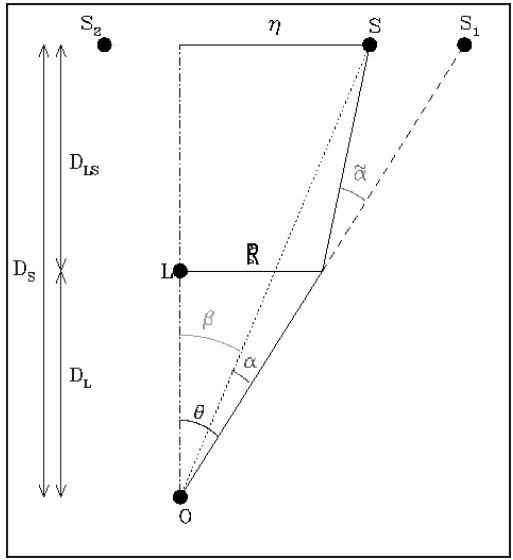}}
\caption{The geometry for a thin lens (adapted from J. Wambsganss).}\label{lensequation}
\end{figure}

The lens equation relates the true position on the sky of a source to the position of its image(s).   We can get the basic idea of
gravitational lensing by considering a point mass, or a
circularly-symmetric mass distribution.  Fig.
\ref{lensequation} shows the geometry of the situation.
The lens equation comes simply by noting that $PS_1 = PS+SS_1$ (where P is the unmarked point directly above L).  If we
denote the observer-source distance by $D_S$, the lens-source
distance by $D_{LS}$, the position on the sky of the image by an
angle $\theta$, and the position on the sky of the unlensed source
by $\beta$, then, for small angles, this translates to
\begin{equation}
D_S \theta = D_S \beta + D_{LS}\tilde\alpha
\end{equation}
and we note that the bend angle is a function of the distance of the
impact parameter of the ray $R$, $\tilde\alpha=\tilde\alpha(R=D_L\theta)$, where
$D_L$ is the distance to the lens. For cosmological applications, the $D_i$ are angular diameter distances.

For convenience, we define the scaled bend angle by
\begin{equation}
\alpha = {D_{LS}\over D_S}\tilde\alpha.
\end{equation}
This gives us the {\em lens
equation}
\begin{equation}
\beta = \theta - \alpha.
\end{equation}

Note that for a given source position $\beta$, this is an implicit
equation for the image position(s) $\theta$, and for general
circularly-symmetric mass distributions, the lens equation cannot be
solved analytically.  There may indeed be more than one solution,
leading to multiple images of the same source. The lens equation is,
however, an explicit equation for the source position given an image
position. This is a useful feature which can be exploited for
ray-tracing simulations of thin-lens systems, when one can ray-trace
backwards from the observer to the sources.

\subsubsection{Point mass lens: Multiple images and Einstein Rings}

The generic possibility of multiple images can be illustrated nicely
by a point mass lens, where the implicit lens equation can be solved
analytically for the image positions.  For a point mass $M$, the
bend angle is $\tilde\alpha(R) = 4GM/(Rc^2) = 4 GM/(D_L \theta c^2)$, so
the lens equation is
\begin{equation}
\beta = \theta - {4 GM \over c^2 \theta}{D_{LS}\over D_L D_S}
\end{equation}
which is a quadratic for $\theta$:
\begin{equation}
\theta^2 - \beta\,\theta - \theta_E^2 = 0
\end{equation}
where we have defined the {\em Einstein Angle}
\begin{equation}
\theta_E \equiv \sqrt{{4GM\over c^2}{D_{LS}\over D_L D_S}}.
\end{equation}
There are evidently two images of the source for a point mass lens,
one either side of the lens, at angles of
\begin{equation}
\theta_\pm = {\beta\over 2} \pm \sqrt{{\beta^2\over 4}+\theta_E^2}.
\label{PointLensSolution}
\end{equation}
The solutions are the
intersections of the line $\alpha(\theta)$ and the
straight line $\theta-\beta$.  For $\theta<0$, the bend is in the
opposite direction, so $\alpha<0$.  We see that, with the point
mass, there are inevitably two solutions for $\beta\ne 0$.

There is a special case when the observer, lens and source are lined
up ($\beta=0$), when one singular solution is the straight-line path,
and the other, at $\theta=\theta_E$, is an {\em Einstein Ring}.
Clearly, from the rotational symmetry of the arrangement, the image
will appear at $\theta_E$ in all directions round the lens.  A
near-perfect Einstein ring is shown in the colour plate section.

Notice that if $|\beta| \gg \theta_E$, then the main image is
perturbed only slightly, at $\beta + \theta_E^2/\beta + O(\beta^{-3})$,
and the second image is very close to the lens at $\theta_-\simeq
-\theta_E^2/\beta$. As we will see shortly, this second image is very
faint.  $\theta_E$ is a useful ballpark angle for determining
whether the deflection of the source rays is significant or not.
Typically, it is about an arcsecond for lensing by large galaxies at
cosmological distances, and microarcseconds for lensing of stars in
nearby galaxies by stars in the Milky Way.

\subsubsection{Magnification and Amplification}

As we have seen, simple lenses alter the positions of the image of
the source, and may indeed produce multiple images.  Another
important, and detectable, effect is that the apparent size and the
brightness of the source will change when its light undergoes a
lensing event.  The gravitational bending of light preserves surface
brightness, 
so the change in apparent size is accompanied by a similar change in brightness.

For the circularly-symmetric lenses, the ratio of the solid angle of
the source and that of the image gives the amplification of an
infinitesimally small source:
\begin{equation}
A = {\theta\over \beta}{d\theta\over d\beta}.
\end{equation}
For a point lens, the amplifications of the two images are obtained
straightforwardly by differentiating equation
(\ref{PointLensSolution}):
\begin{equation}
A_{\pm} = {1\over 2}\left(1\pm
\frac{\beta^2+2\theta_E^2}{\beta\sqrt{\beta^2+4\theta_E^2}}\right).
\label{Apm}
\end{equation}
Note that the amplification can be negative.  This corresponds to an
image which is flipped with respect to the source.

In some cases, such as microlensing, where the two images are
unresolved, one can only measure the total amplification,
\begin{equation}
A_{\rm total} =
|A_{+}|+|A_{-}|=\frac{\beta^2+2\theta_E^2}{\beta\sqrt{\beta^2+4\theta_E^2}}
\label{Atot}
\end{equation} and we note that the difference of the
amplifications of images by a point lens is unity.  In the limit
$|\beta|\gg \theta_E$, $A_+\to 1+(\theta_E/\beta)^4$ and $A_-\to
-(\theta_E/\beta)^4$, so the inner image is extremely faint in this
limit.

\subsection{General thin lens mass distributions}

For a general distribution of mass, the bend angle is a 2D vector on
the sky, $\balpha$.  The lens equation then generalises
naturally to a vector equation
\begin{equation}
\bbeta = \vth - \balpha.\label{VectorLens}
\end{equation}
For a thin lens with surface density $\Sigma(\vth\,')$, the bend
due at $\vth$ to a small element of mass
$dM=\Sigma(\vth\,')\,d^2\vth\,'$ is evidently in the direction
$\vth-\vth'$, and has a magnitude
$4GdM/(c^2|\vth-\vth\,'|)$, so the total bend angle is
\begin{equation}
\balpha(\vth) = {4GD_L D_{LS}\over D_S c^2}\int d^2\vth\,'
{\Sigma(\vth\,')(\vth-\vth\,')\over
|\vth-\vth\,'|^2}.
\end{equation}
The bend angle can be expressed as the (2D) gradient of the (thin
lens) {\em lensing potential},  $\balpha = \nabla\phi$, where
\begin{equation}
\phi(\vth) = {4GD_L D_{LS}\over c^2 D_S}\int d^2\vth\,'
\Sigma(\vth\,')\ln(|\vth-\vth\,'|)\qquad {\rm (thin\ lens)}.\label{Green}
\end{equation}
This potential satisfies the 2D Poisson equation
\begin{equation}
\nabla^2\phi(\vth) = 2\kappa(\vth)
\end{equation}
where $\nabla$ is a 2D gradient, with respect to angle.
$\kappa$ is the {\em convergence} $\kappa(\vth) \equiv \Sigma(\vth)/\Sigma_{\rm crit}$, with
\begin{equation}
\Sigma_{crit} \equiv \frac{c^2 D_S}{4\pi G D_L D_{LS}}
\end{equation}
being the {\em critical surface density}.
The 2D Poisson equation is extremely useful, as it allows us to estimate the surface mass density of an intervening lens from lensing measurements, as we shall see later.

\subsubsection{Convergence, magnification and shear for general thin lenses}

\begin{figure}
\sidecaption
\includegraphics[height=6cm]{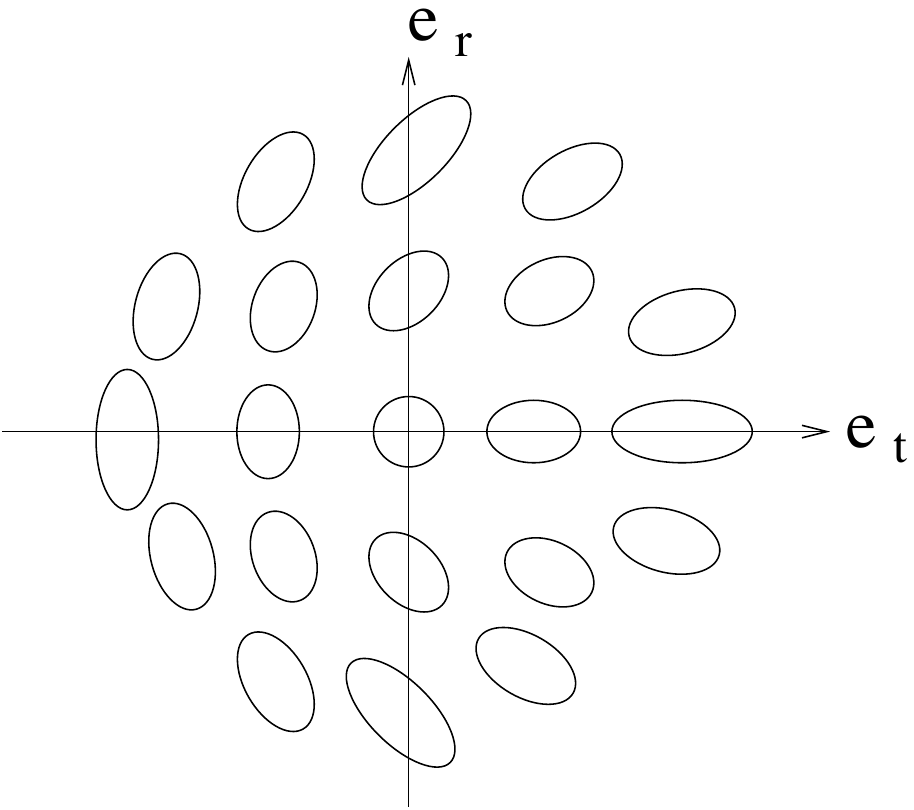}
\caption{The effect of shear distortions on a circular source.  In the notation of the current paper, $e_t$ and $e_r$ are the real and imaginary parts of the ellipticity (or shear).  From \cite{VWM}.}
\label{shearfig}
\end{figure}

For a general distribution of mass within a thin lens, the
magnification and distortion of an infinitesimal source are given by
the transformation matrix from the source position $\bbeta$ to the
image position(s) $\vth$.  From the vector lens equation
(\ref{VectorLens}).  The (inverse) amplification matrix is
\begin{equation}
A_{ij} \equiv \frac{\partial \beta_i}{\partial \theta_j} = \delta_{ij}-\phi_{ij},\label{defA}
\end{equation}
where $\phi_{ij}\equiv \partial^2\phi/\partial\theta_i\partial\theta_j$.  We see that $A$ is symmetric, and it can be decomposed into an isotropic expansion term, and a shear. A
general amplification matrix also includes a rotation term (the
final degree of freedom being the rotation angle), but we see that
weak lensing doesn't introduce rotation of the image, and has only
3 degrees of freedom, rather than the four possible in a $2 \times
2$ matrix.  We decompose the amplification matrix as follows:
\begin{equation}
A_{ij} = \left( \begin{array}{cc}
1-\kappa & 0 \\
0 & 1-\kappa \end{array} \right) + \left( \begin{array}{cc}
-\gamma_1 & -\gamma_2 \\
-\gamma_2 & \gamma_1 \end{array} \right)
\end{equation}
where $\kappa$ is called the {\em convergence} and
\begin{equation}
\gamma=\gamma_1+ i \gamma_2
\end{equation}
is the {\em complex shear}.  For weak lensing, both $|\kappa|$ and $|\gamma_i|$ are $\ll 1$.  A
non-zero $\kappa$ represents an isotropic expansion or contraction of a source; $\gamma_1>0$ represents an elongation of the image
along the $x$-axis and contraction along $y$.  $\gamma_1<0$ stretches along $y$ and contracts along $x$. $\gamma_2\ne 0$
represents stretching along $x=\pm y$ directions.  The effects are shown in Fig.\ref{shearfig}.

Making the decomposition, we find that
\begin{eqnarray}
\kappa &=& {1\over 2}\left(\phi_{11}+\phi_{22}\right)\\\nonumber
\gamma_1 &=&{1\over 2}\left(\phi_{11}-\phi_{22}\right)\equiv D_1\phi\\\nonumber
\gamma_2 &=& \phi_{12}\equiv D_2\phi.\label{shearkappa}
\end{eqnarray}
which defines the two operators $D_{1,2}$,
and it is straightforward to prove that
\begin{equation}
D_i D_i = (\nabla^2)^2\label{DD},
\end{equation}
where the summation is over $i=1,2$.

Note that $\kappa>0$ corresponds to {\em magnification} of the
image. Lensing preserves surface brightness, so this also amounts
to {\em amplification} of the source flux. The magnification is
\begin{equation}
A = \frac{1}{\det A_{ij}} = \frac{1}{(1-\kappa)^2-|\gamma|^2}.
\end{equation}
We see that we may have infinite amplifications if $\kappa\ge 1$.  Such effects apply only for infinitesimal sources, and places in the source plane which lead to infinite magnifications are referred to as {\em caustics}, and lead to highly distorted images along lines called {\em critical lines} in the lens plane.  The giant arcs visible in images of some rich clusters lie on or close to critical lines.  For cosmic shear, due to the general inhomogeneities along the line-of-sight, $\kappa$ and $|\gamma|$ are typically 0.01, and the lensing is weak.

It is worth noting that the amplification matrix may be written
\begin{equation}
A_{ij} = (1-\kappa)\left( \begin{array}{cc}
1-g_1 & -g_2 \\
-g_2 & 1+g_1 \end{array} \right)
\end{equation}
where $g\equiv \gamma/(1-\kappa)$ is called the {\em reduced shear}.
Since the $1-\kappa$ multiplier affects only the overall size (and
hence brightness) of the source, but not its shape, we see that shear measurements can
determine only the reduced shear, and not the shear itself.  For
weak lensing, $\kappa\ll 1$, so the two coincide to linear order.

Note that a rotation of $\pi/4$ in the coordinate system changes a real shear into an imaginary shear - i.e. the complex shear rotates by $\pi/2$, twice the angle of rotation of the coordinate system.  This behaviour is characteristic of a {\em spin-weight 2} field, and is encountered also in microwave background polarisation and gravitational wave fields.

The shear field is in principle observable, so let us see how it can be used to estimate the surface mass density of a lens.

\subsubsection{Estimating shear}

The estimation of shear from galaxy image data is a complex business, and I will give no more than a highly simplified sketch.  For more details, see reviews such as \cite{Munshi08,VWM, HJ08}.   One way to estimate shear is to measure the {\em complex ellipticity} $e$ of a galaxy, which can be defined in terms of moments \cite{KSB} even if the galaxy image is not elliptical in shape. Fig. \ref{shearfig} shows how simple shapes map onto ellipticity.  In the limit of weak distortions, the observed ellipticity is
\begin{equation}
e = \frac{e^s + \gamma}{1+\gamma^*e^s}
\end{equation}
where $e^s$ is the undistorted source ellipticity.  The source ellipticity has a dispersion $\langle |e^s|^2 \rangle \simeq 0.3$, whereas $\gamma$ itself is usually much smaller, 0.01-0.1.   If we average over some sources,
\begin{equation}
\langle e \rangle = \gamma
\end{equation}
so we can use the average ellipticity as an estimator of shear.

Two points to note:

1. The estimator is noisy, dominated by the intrinsic dispersion of $e$, so $\gamma$ has a variance $\sim \langle |e^s|^2\rangle/N$ if $N$ sources are averaged.

2. In these notes the formulae will often refer to $\gamma$; it should be noted that in reality it will be $e$, a noisy estimate of $\gamma$, which is used in practice.

The main practical difficulty of lensing experiments is that the
atmosphere and telescope affect the shape of the images.  These
modifications to the shape may arise due to such things as the
point spread function, or poor tracking of the telescope.  The
former needs to be treated with great care.  Stars (whose images
should be round) can be used to remove image distortions to very
high accuracy, although a possibly fundamental limitation may
arise because of the finite number of stars in an image.
Interpolation of the anisotropy of the PSF needs to be done
carefully, and examples of how this can be done in an optimal way
are given in \cite{VWMH}.    Bayesian methods are beginning to be used, with {\em lensfit} \cite{Miller07,Kitching08b} showing very promising results.

\section{Dark Matter}

Theoretical work with numerical simulations indicates that in the absence of the effects of baryons, virialised Dark Matter haloes should follow a uniform `NFW' profile, $\rho(r) = \rho_s (r_s/r)(1+r/r_s)^{-2}$ \cite{NFW}, if the Dark Matter is cold (CDM).  Simulations also predict how the physical size of the clusters should depend on mass, characterised by the concentration index $c_s \equiv r_{vir}/r_s$, where $r_{vir}$ is defined as the radius within which the mean density is 200 times the background density.  Roughly $c_s \propto M^{-0.1}$. This can be tested, by measuring the shear signal and stacking the results from many haloes to increase signal-to-noise. 
\subsection{2D mass surface density reconstruction}

We wish to take a map of estimated shear (ellipticities) and estimate the surface mass density of the intervening lens system.  For simplicity here we assume the sources are at the same distance.  The classic way to do this was given by \cite{KaiserSquires}.

In some respects it is easiest to work in Fourier space.  Expanding
\begin{equation}
\kappa_{\bf k} \equiv \int d^2\theta \,\kappa(\vth) \,\exp(i{\bf k}.\vth)
\end{equation}
etc, then evidently
\begin{eqnarray}
\kappa_{\bf k} &=& -\frac{1}{2}k^2\phi_{\bf k} \\\nonumber
\gamma_{1{\bf k}} &=& -\frac{1}{2}(k_1^2-k_2^2)\phi_{\bf k} \\\nonumber
\gamma_{2{\bf k}} &=& -k_1 k_2\phi_{\bf k} \\\nonumber
\end{eqnarray}
where $k^2 = k_1^2+k_2^2=|{\bf k}|^2$.

We see that the following are estimators of $\kappa_{\bf k}$:
\begin{equation}
\gamma_{1{\bf k}}\frac{k^2}{k_1^2-k_2^2}; \qquad
\gamma_{2{\bf k}}\frac{k^2}{2 k_1 k_2}. 
\end{equation}
In practice, we use the shapes of galaxies to estimate shear, so our estimates of
$\gamma_{i{\bf k}}$ are dominated by noise in the form of scatter in the intrinsic shapes of galaxies.  Thus the variance of each of these estimators is determined by the variances in $|e_{i{\bf k}}|^2$.  The variance of the first estimator is therefore proportional to $k^4/(k_1^2-k_2^2)^2$, and the second to $k^4/(2k_1 k_2)^2$.  The optimal estimator for $\kappa_{\bf k}$ is given by the standard inverse variance weighting, giving
\begin{equation}
\kappa_{\bf k} = \left(\frac{k_1^2-k_2^2}{k^2}\right)\gamma_{1{\bf k}}+\left(\frac{2 k_1 k_2}{k^2}\right)\gamma_{2{\bf k}}.
\end{equation}
Since this is the sum of two terms, each of which is a multiplication in ${\bf k}$ space, it represents a convolution in real space. (An exercise for the enthusiastic reader is to Fourier transform to find the real-space convolution). In fact it is easier to find the convolution by noting that
\begin{eqnarray}
\gamma_i &=& D_i \phi\\\nonumber
&=&2D_i\nabla^{-2}\kappa\\\nonumber
\Rightarrow \kappa &=& 2 D_i\nabla^{-2}\gamma_i,
\end{eqnarray}
where the last line follows from equation \ref{DD}.  Now we know the solution to the 2D Poisson equation - it is given in equation \ref{Green}:
\begin{equation}
\nabla^{-2}\gamma_i(\vth) = \frac{1}{2\pi}\int d^2\vth' \gamma_i(\vth')\ln|\vth'-\vth|.
\end{equation}
Upon differentiation with $D_i$ and summation, we find
\begin{equation}
\kappa(\vth) = \frac{2}{\pi}\,\int d^2\vth' \frac{[\gamma_1(\vth') \cos(2\psi) + \gamma_2(\vth') \sin(2\psi)]}{|\vth'-\vth|^2}\label{KS}
\end{equation}
where $\psi$ is the angle between $\vth$ and $\vth'$.

It is quite tempting to work with equation \ref{KS} in its discrete form, replacing the integral by a sum over shear estimates at the galaxy locations $g$, i.e. considering the following estimator for $\kappa$:
\begin{equation}
\hat\kappa(\vth) = \frac{1}{\bar n} \sum_g \frac{[\gamma_1(\vth_g) \cos(2\psi) + \gamma_2(\vth_g) \sin(2\psi)]}{|\vth_g-\vth|^2}\label{KSD}
\end{equation}
where $\bar n$ is the mean surface density of sources.
This estimator would, however, be a mistake.  It is unbiased, but it has the awkward and undesirable property of having infinite noise.  This can be seen as follows.

The fourier transform of equation \ref{KSD} is
\begin{equation}
\hat\kappa_{\bf k} = \frac{1}{\bar n}\left[\left(\frac{k_1^2-k_2^2}{k^2}\right)\sum_g\gamma_1(\vth_g)\exp(i{\bf k}.\vth_g)+\left(\frac{2 k_1 k_2}{k^2}\right)\sum_g \gamma_{2}(\vth)\exp(i{\bf k}.\vth_g)\right].\label{KappaK}
\end{equation}
If we assume (not quite correctly - see a later discussion on intrinsic alignments; but this will do here) that the galaxy shapes are uncorrelated, then
\begin{equation}
\left\langle\hat\kappa_{\bf k}\hat\kappa^*_{{\bf k}}\right\rangle = \frac{\langle |e^2| \rangle}{2\bar n}\label{KK}
\end{equation}
where $\langle |e^2| \rangle$ is the dispersion in galaxy ellipticities.  Note that as usual, the estimate of the shear is dominated by the intrinsic ellipticity.

If we use Parseval's theorem, we see that the variance in the recovered convergence, being the integral of equation \ref{KK} over ${\bf k}$, diverges.

The solution is to smooth the field, which can be done at any stage - either smoothing the estimated shears (or averaging them in cells), or by working in Fourier space and introducing a filter, or even by smoothing the estimated noisy convergence field.  The amount of smoothing required can prevent high-resolution maps of all but the richest clusters, but the advent of high surface densities with space telescopes has improved this situation considerably.

Finally, we note that the estimate of the zero-wavenumber coefficient $\kappa_{\bf 0}$ is undetermined by equation \ref{KappaK}.  We cannot determine the mean surface density this way.  This feature is an example of the so-called `mass-sheet degeneracy'.  Other measurements (amplification/magnification, or by observing sources at different distances behind the lens) can alleviate this problem, or one can assume that $\kappa \rightarrow 0$ as one goes far from the lens centre.
\begin{figure}
\sidecaption
\includegraphics[height=7cm]{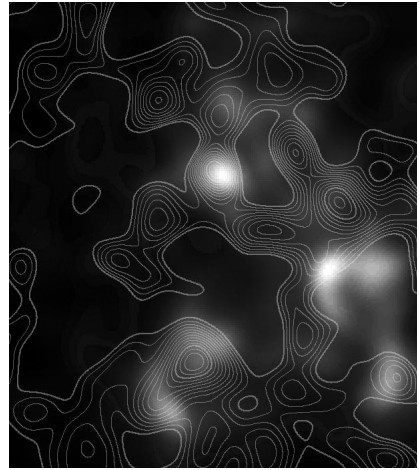}
\caption{2D reconstruction of surface matter density from the COMBO17 A901 study (contours; \cite{Gray04}) superimposed on galaxy surface density.}
\label{A901}
\end{figure}
\subsection{Testing the Navarro-Frenk-White profile of CDM}

\begin{figure}
\sidecaption
\includegraphics[width=7cm]{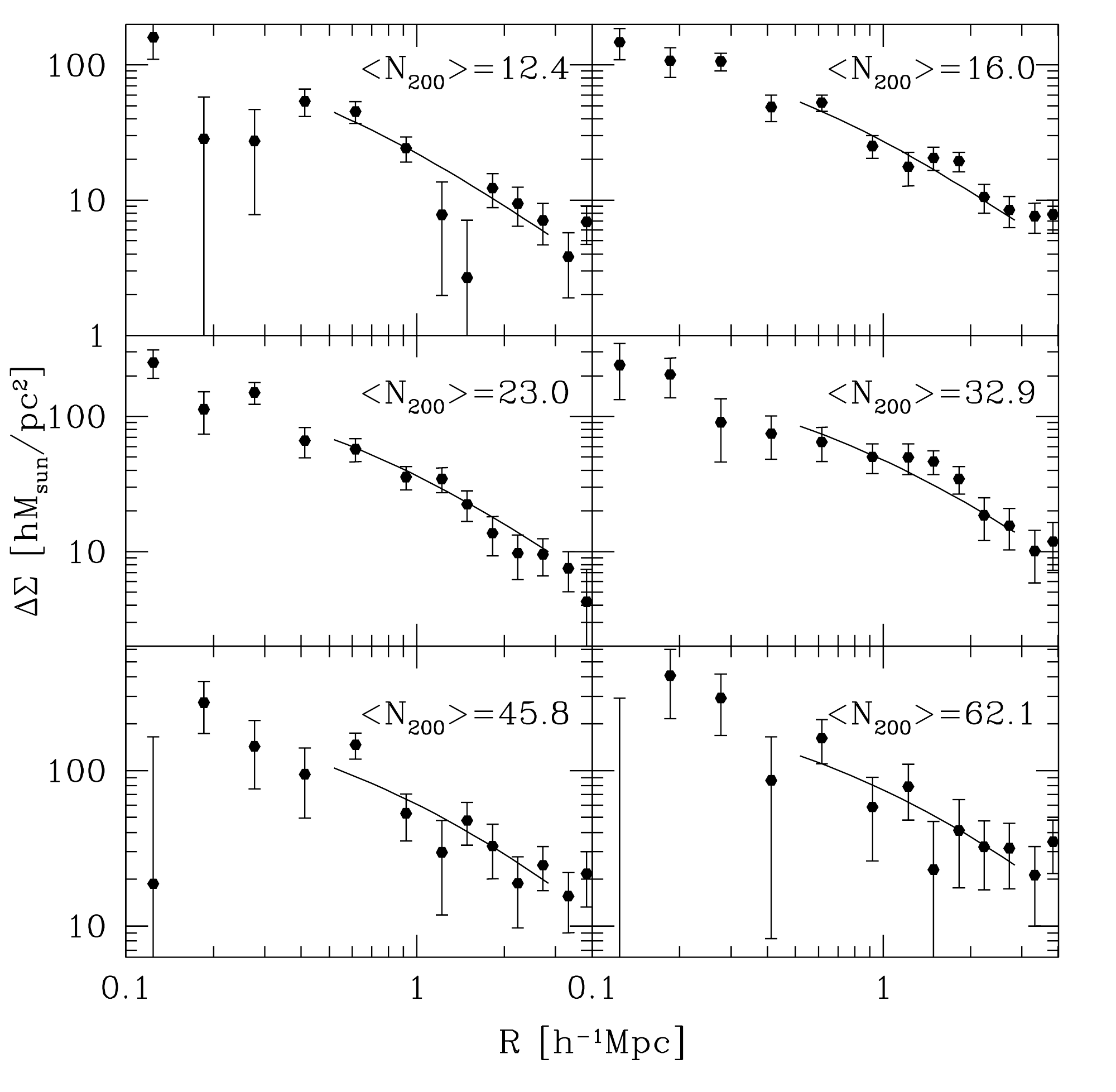}
\caption{Excess surface density from stacked galaxy clusters from the SDSS survey, with best-fitting NFW profiles.  $N_{200}$ is a measure of the richness of the clusters.  From \cite{Mandelbaum08}.
\label{SDSSNFW}}
\end{figure}

\begin{figure}
\sidecaption
\includegraphics[width=2.787in]{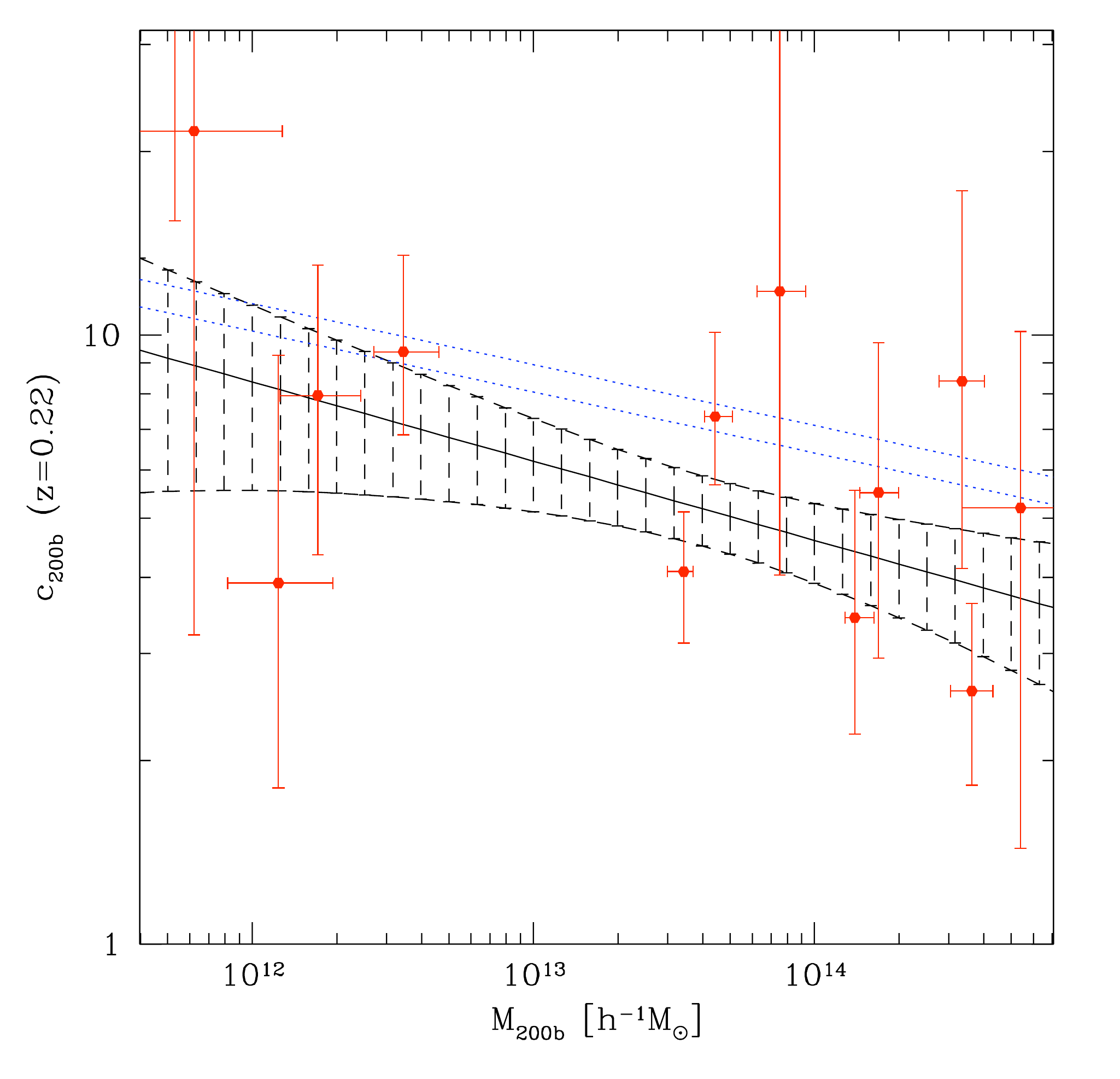}
\caption{Concentration indices from SDSS clusters as a function of mass, compared with simulation (dotted, for different cosmologies).  Dashed regions show range assuming a power-law $c_s-M$ relation.  From \cite{Mandelbaum08}.
\label{SDSSCM}}
\end{figure}

Fig. \ref{SDSSNFW} shows the average radial surface density profiles for clusters identified in the Sloan Digital Sky Survey (SDSS), grouped by number of cluster galaxies, and NFW fits superimposed  \cite{Mandelbaum08}.  
Fig.\ref{SDSSCM} shows that the observed concentration indices are close to the theoretical predictions, but some tension exists.    Broadly, weak lensing data on clusters therefore supports the CDM model.  We will consider 3D mapping later, but unfortunately the limited accuracy of photo-zs ($\sim 0.03 \simeq 100 h^{-1}$ Mpc typically) means the 3D mass map is smoothed heavily in the radial direction, and this limits the usefulness of 3D mapping for testing the NFW profile.  

A more radical test of theory has been performed with the Bullet Cluster \cite{Clowe04}, actually a pair of clusters which have recently passed through each other.  There are two clear peaks in the surface density of galaxies, and X-ray emission from hot shocked gas in between.  In the standard cosmological model, this makes perfect sense, as the galaxies in the clusters are essentially collisionless.  If the (dominant) Dark Matter is also collisionless, then we would expect to see surface mass concentrations at the locations of the optical galaxy clusters, and this is exactly what is observed.  In MOND or TeVeS models without Dark Matter, one would expect the surface mass density to peak
where the dominant baryon component is - the X-ray gas.  This is not seen.  A caveat is that it is not quite the surface density which is observed, rather the {\em convergence}, which is related to the distortion pattern of the galaxy images, and which is proportional to surface density in General Relativity (GR), but not in MOND/TeVeS.  However, no satisfactory explanation of the bullet cluster has been demonstrated without Dark Matter.

\section{Cosmological lensing}

Gravitational lensing is strong if the distortions to the images are
substantial, and weak if the distortions are small. Weak lensing of
background images can be used to learn about the mass distribution
in individual objects, such as galaxy clusters, but we
concentrate now on weak lensing on a cosmological scale, which is
usually analysed in a statistical way.  Reviews of this area
include \cite{Munshi08,HJ08,BS,Schneider,VWM}.

The basic effect of weak lensing is for the clumpy matter
distribution to perturb slightly the trajectories of photon paths.
By considering how nearby light paths are perturbed, one finds
that the shapes of distant objects are changed slightly.
Associated with the size change is a change in the brightness of
the source.   Size and magnitude changes can, in principle, be
used to constrain the properties of the matter distribution along
the line-of-sight (and cosmological parameters as well), but it is
the change in shape of background sources which has almost
exclusively been used in cosmological weak lensing studies.   The
reason for this is simply that the signal-to-noise is better.
These notes will concentrate on shear (=shape changes), but the
magnification and amplification of sources can also be used and
will probably be used in future when the surveys are larger.  The
great promise of lensing is that it acts as a direct probe of the
matter distribution (whether dark or not), and avoids the use of
objects which are assumed to trace the mass distribution in some
way, such as galaxies in large-scale structure studies.
Theoretically, lensing is very appealing, as the physics is very
simple,  and very robust, direct connections can be made between
weak lensing observables and the statistical properties of the
matter distribution.  These statistical properties are dependent
on cosmological parameters in a known way, so weak lensing can be
employed as a cosmological tool.  The main uncertainties in
lensing are observational - it is very challenging to make images
of the necessary quality.  In this section, we concentrate here on the weak effects
on a cosmological scale of the non-uniform distribution of matter
all the way between the source and observer, an effect often
referred to as {\em cosmic shear}.

\subsection{Distortion of light bundles}

The distortion of a light bundle has to be treated with General
Relativity, but if one is prepared to accept one modification to
Newtonian physics, one can do without GR.

In an expanding Universe, it is usual to define a {\em comoving
coordinate} {\bf x}, such that `fundamental observers' retain the same
coordinate.  Fundamental observers are characterised by the property
of seeing the Universe as isotropic; the Earth is not (quite) a
fundamental observer, as from here the Cosmic Microwave Background
looks slightly anisotropic. The equation of motion for the
transverse coordinates (about some fiducial direction) of a photon
in a flat Universe is
\begin{equation}
{d^2x_i\over d \eta^2} = -{2\over c^2}\,{\partial \Phi\over
\partial x_i}. \qquad i=1,2\label{eom}
\end{equation}
We assume a flat Universe throughout, but the generalisation to non-flat Universes is straightforward (there is an extra term in the equation above, and some $r$ symbols need to be changed to an angular diameter distance).

The equation of motion can be derived using General Relativity (see Appendix for
details).  We will use $x_i, i=1,2$ for coordinates transverse to
the line-of-sight, and $r$ to indicate the radial coordinate. $\eta$
is the conformal time, related to the coordinate $t$ by $d\eta = c
dt/R(t)$ and $R(t)$ is the cosmic scale factor, equal to $R_0$ at
the present time. $\Phi(x_i,r)$ is the peculiar gravitational
potential, related to the matter overdensity field $\delta \equiv
\delta\rho/\rho$ by Poisson's equation
\begin{equation}
\nabla^2_{3D} \Phi = {3H_0^2 \Omega_m\over 2 a(t)}\delta
\label{Poisson}
\end{equation}
where $H_0$ is the Hubble constant, $\Omega_m$ is the present matter
density parameter, and $a(t)= R(t)/R_0 = (1+z)^{-1}$, where $z$ is
redshift.

The equation of motion is derived in the Appendix from General
Relativity, in a (nearly flat) metric given in the Newtonian gauge by
\begin{equation}
ds^2 =
(1+2\Phi/c^2)c^2 dt^2 - (1-2\Phi/c^2)R^2(t)(dr^2 + r^2 d\theta^2 +
r^2 \sin^2\theta d\varphi^2).
\end{equation}
From a Newtonian point-of-view,
equation \ref{eom} is understandable if we note that time is
replaced by $\eta$ (which arises because we are using comoving
coordinates), and there is a factor 2 which does not appear in
Newtonian physics. This same factor of two gives rise to the famous
result that in GR the angle of light bending round the Sun is double
that of Newtonian theory.

The coordinates ${\bf x}_i$ are related to the (small) angles of the
photon to the fiducial direction $\vth=(\theta_x,\theta_y)$ by
${\bf x}_i = r \vth_i$.

\subsection{Lensing potential}

The solution to (\ref{eom}) is obtained by first noting that the
zero-order ray has $ds^2=0 \Rightarrow dr=-d\eta$, where we take
the negative root because the light ray is incoming. Integrating
twice, and reversing the order of integration gives
\begin{equation}
x_i = r\theta_i - {2\over c^2}\int_0^r dr' {\partial \Phi\over
\partial x'_i} (r-r').
\end{equation}
We now perform a Taylor expansion of $\partial \Phi/
\partial x'_i$ , and find the deviation of
two nearby light rays is
\begin{equation}
\Delta x_i = r\Delta \theta_i - {2\over c^2} \Delta \theta_j
\int_0^r dr' r'(r-r') {\partial^2 \Phi\over \partial x'_i \partial
x'_j}
\end{equation}
which we may write as
\begin{equation}
\Delta x_i = r \Delta \theta_j(\delta_{ij} - \phi_{ij})
\end{equation}
where $\delta_{ij}$ is the Kronecker delta ($i=1,2$) and we define
\begin{equation}
\phi_{ij}({\bf r}) \equiv {2\over c^2}\int_0^r dr' {(r-r')\over r r'}
{\partial^2 \Phi({\bf r}')\over
\partial \theta_i \partial \theta_j}.
\end{equation}
The integral is understood to be along a radial line (i.e. ${\bf r}
\parallel {\bf r}'$); this is the {\em Born approximation}, which is
a very good approximation for weak lensing
\cite{Bernardeau,Schneider98,VW2002}.  In reality the light path is
not quite radial.

It is convenient to introduce the {\em (cosmological) lensing potential}, which
controls the distortion of the ray bundle:
\begin{equation}
\phi({\bf r}) \equiv {2\over c^2}\int_0^r dr' {(r-r')\over r r'}
\Phi({\bf r}')\qquad {\rm (cosmological;\ flat\ universe)}
\end{equation}
Note that $\phi_{ij}({\bf r}) =
\partial^2 \phi({\bf r})/\partial \theta_i
\partial \theta_j$.
So, remarkably, we can describe the distortion of an image as it passes
through a clumpy universe in a rather simple way.   The shear and convergence of a source image is obtained
from the potential in the same way as the thin lens case (equation \ref{shearkappa}), although the relationship between
the convergence and the foreground density is more complicated, as we see next.

\subsubsection{Relationship to matter density field}

The gravitational potential $\Phi$ is related to the matter
overdensity field $\delta \equiv \delta\rho/\rho$ by Poisson's
equation (\ref{Poisson}).  The convergence is then
\begin{equation}
\kappa({\bf r}) = {3 H_0^2 \Omega_m\over 2 c^2}\int_0^r dr'
{r'(r-r')\over r} {\delta({\bf r}')\over a(r')}
\end{equation}
Note that there is an extra term $\partial^2\Phi/\partial r'^2$ in
$\nabla^2_{3D}$ which integrates to zero to the order to which we are
working.

\subsubsection{Averaging over a distribution of sources}

If we consider the distortion averaged over a distribution of
sources with a radial distribution $p(r)$ (normalised such that
$\int dr\, p(r)=1$), the average distortion is again obtained by
reversing the order of integration:
\begin{equation}
\Delta x_i = r\Delta\theta_j\left(\delta_{ij}-{2\over c^2}\int_0^r
{dr'\over r'}\,g(r') {\partial^2 \Phi({\bf r}')\over
\partial \theta_i \partial \theta_j}\right)
\end{equation}
where
\begin{equation}
g(r) \equiv \int_r^\infty dr'\, p(r') {r'-r\over r'}.
\end{equation}
In order to estimate $p(r)$, surveys began to estimate distances
to source galaxies using photometric redshifts.  This has opened
up the prospect of a full 3D analysis of the shear field, which we
will discuss briefly later in this article.

\subsubsection{Convergence power spectrum and Shear correlation
function}

The average shear is zero, so the most common statistics to use
for cosmology are two-point statistics, quadratic in the shear.
These may be in `configuration' (`real') space, or in transform
space (using Fourier coefficients or similar).  I will focus on
two quadratic measures, the convergence power spectrum and the
shear-shear correlation function.

To find the expectation value of a quadratic quantity, it is
convenient to make use of the matter density power spectrum,
$P(k)$, defined by the following relation between the overdensity
Fourier coefficients:
\begin{equation}
\langle \delta_{\bf k} \delta^*_{{\bf k}'}\rangle = (2\pi)^3
\delta^D({\bf k}-{\bf k}') P(k),
\label{power}
\end{equation}
where $\delta^D$ is the Dirac delta function.  $P(k)$ is evolving,
so we write it as $P(k;r)$ in future, where $r$ and $t$ are related
through the lookback time.  (This $r$-dependence may look strange;
there is a subtlety: (\ref{power}) holds if the field is homogeneous
and isotropic, which the field on the past light cone is not, since
it evolves.  In the radial integrals, one has to consider the
homogeneous field at the same cosmic time as the time of emission of
the source). The trick is to get the desired quadratic quantity into
a form which includes $P(k;r)$.

For the convergence power spectrum, we first transform the
convergence in a 2D Fourier transform on the sky, where $\ell$ is
a 2D dimensionless wavenumber:
\begin{eqnarray}
\kappa_\ell &=& \int d^2\vth\, \kappa(\vth) e^{-i\bell.\vth}\\
 &=&
A\int_0^\infty dr\, r\, {g(r)\over a(r)}\int d^2\vth\,
\delta(r\vth,r)e^{-i\bell.\vth}\label{kappa2D}
\end{eqnarray}
where $A\equiv 3H_0^2\Omega_m/2c^2$. We expand the overdensity
field in a Fourier transform,
\begin{equation}
\delta(r\vth,r) = \int {d^3{\bf k}\over (2\pi)^3}\, \delta_{\bf k} e^{i
k_\parallel r} e^{i{\bf k}_\perp.r\vth}
\end{equation}
and substitute into (\ref{kappa2D}).  We form the quantity
$\langle \kappa_\ell \kappa^*_{\ell'}\rangle$, which, by analogy
with (\ref{power}), is related to the (2D) {\em convergence power
spectrum} by
\begin{equation}
\langle \kappa_\ell \kappa^*_{\ell'}\rangle = (2\pi)^2
\delta^D(\bell-\bell') P_\kappa(|\ell|).
\end{equation}
Straightforwardly,
\begin{eqnarray}
\langle \kappa_\ell \kappa^*_{\ell'}\rangle &=& A^2 \int_0^\infty
dr\, G(r) \int_0^\infty dr'\, G(r') \int d^2\vth d^2\vth' {d^3
{\bf k}\over (2\pi)^3} {d^3 {\bf k}'\over (2\pi)^3} \\\nonumber & &
\langle \delta_{\bf k} \delta_{{\bf k}'}^* \rangle \exp(ik_\parallel r - i
{k_\parallel}'r') \exp(i{\bf k}_\perp.\vth-i{\bf k}'_\perp.\vth')\exp(-i\bell.\vth+i\bell'.\vth')
\end{eqnarray}
where $G(r)\equiv rg(r)/a(r)$. Using (\ref{power}) we remove the
${\bf k}'$ integration, introducing the power spectrum
$P(k)=P(\sqrt{k_\parallel^2+|{\bf k}_\perp|^2})$.  For small-angle
surveys, most of the signal comes from short wavelengths, and the
$k_\parallel$ is negligible, so $P(k)\simeq P(|{\bf k}_\perp|)$.  The
only $k_\parallel$ term remaining is the exponential, which
integrates to $(2\pi)\delta^D(r-r')$.  The integrals over $\vth$
and $\vth'$ give $(2\pi)^2\delta^D(\bell-r{\bf k}_\perp)$ and
$(2\pi)^2\delta^D(\bell'-r{{\bf k}_\perp}')$ respectively, so the
whole lot simplifies to give the convergence power spectrum as
\begin{equation}
P_\kappa(\ell) = \left({3H_0^2 \Omega_m\over 2 c^2}\right)^2
\int_0^\infty dr\,\left[{g(r)\over a(r)}\right]^2 P(\ell/r;r).
\end{equation}
An exercise for the reader is to show that the power spectrum for the complex shear field
$\gamma$ is the same: $P_\gamma = P_\kappa$. The shear correlation
function, for points separated by an angle $\theta$ is
\begin{eqnarray}
\langle \gamma \gamma^*\rangle_\theta &=& \int {d^2\ell\over
(2\pi)^2}\, P_\gamma(\ell)\, e^{i\bell.\vth}\\\nonumber
 &=& \int {\ell d\ell\over (2\pi)^2}\, P_\kappa(\ell)
 e^{i\ell\theta\cos\varphi} d\varphi\\\nonumber
 &=& \int {d\ell\over 2\pi} \ell P_\kappa(\ell) J_0(\ell\theta)
 \label{gammacorrn}
 \end{eqnarray}
where we have used polar coordinates, with $\varphi$ the angle
between $\ell$ and $\vth$, and we have exploited the isotropy
($P_\kappa$ depends only on the modulus of $\ell$). $J_0$ is a
Bessel function.

Other quadratic quantities (examples are shear variances on
different scales, Aperture Mass (squared)) can be written
similarly as integrals over the power spectrum, with different
kernel functions.

\subsection{Matter power spectrum}

\begin{figure}
\sidecaption
\includegraphics[height=8cm]{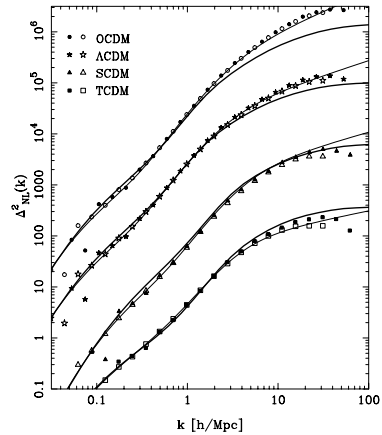}
%
%
\caption{The nonlinear power spectrum from numerical simulations,
along with fitting functions (from \cite{Smith}.}
\label{Smithetal}       
\end{figure}

As we have seen, the two-point statistics of the shear and
convergence fields depend on the power spectrum of the matter,
$P(k;t)$.  The power spectrum grows in a simple way when the
perturbations in the overdensity are small, $|\delta|\ll 1$, when
the power spectrum grows in amplitude whilst keeping the same
shape as a function of $k$.  However, gravitational lensing can
still be weak, even if the overdensity field is nonlinear.
Poisson's equation still holds provided we are in the weak-field
limit as far as General Relativity is concerned, and this
essentially always holds for cases of practical interest.  In
order to get as much statistical power out of lensing, one must
probe the nonlinear regime, so it is necessary for parameter
estimation to know how the power spectrum grows.  Through the
extensive use of numerical simulations, the growth of dark matter
clustering is well understood down to quite small scales, where
uncertainties in modelling, or uncertain physics, such as the
influence of baryons on the dark matter \cite{White}, make the
predictions unreliable.  Accurate fits for the nonlinear power
spectrum have been found \cite{Smith} up to $k > 10 h Mpc^{-1}$,
which is far beyond the linear/nonlinear transition $k \sim 0.2h
Mpc^{-1}$.  Fig. \ref{Smithetal} shows fits for a number of CDM
models.
For precision use, one must make sure that the statistics do not
have substantial contributions from the high-$k$ end where the
nonlinear power spectrum is uncertain.  This can be explored by
looking at the kernel functions implicit in the quantities such as
the shear correlation function (\ref{gammacorrn}).

\subsection{Intrinsic alignments}

The main signature of weak lensing is a small alignment of the
images, at the level of a correlation of ellipticities of $\sim
10^{-4}$.  One might be concerned that physical processes might
also induce an alignment of the galaxies themselves.  In the
traditional lensing observations, the distances of individual
galaxies are ignored, and one simply uses the alignment on the sky
of galaxies, and one might hope that the galaxies will typically
be at such large separations along the line of sight that any
physical interactions would be rare and can be ignored.  However,
the lensing signal is very small, so the assumption that intrinsic
alignment effects are sufficiently small needs to be tested.  This
was first done in a series of papers by a number of groups in
2000-1 e.g.\cite{HRH,CM,Crittenden01,CKB}, and the answer is that
the effects may not be negligible.  The contamination by intrinsic
alignments is highly depth-dependent.  This is easy to see, since
at fixed angular separation, galaxies in a shallow survey will be
physically closer together in space, and hence more likely to
experience tidal interactions which might align the galaxies.  In
addition to this, the shallower the survey, the smaller the
lensing signal.  In a pioneering study, the alignments of nearby
galaxies in the SuperCOSMOS survey were investigated \cite{Brown}.
This survey is so shallow (median redshift $\sim 0.1$) that the
expected lensing signal is tiny.  A non-zero alignment was found,
which agrees with at least some of the theoretical estimates of
the effect.  The main exception is the numerical study of Jing
\cite{Jing}, which predicts a contamination so high that it could
dominate even deep surveys. For deep surveys, the effect, which is sometimes called the `II correlation', is
expected to be rather small, but if one wants to use weak lensing
as a probe of subtle effects such as the effects of altering the
equation of state of dark energy, then one has to do something.
There are essentially two options - either one tries to calculate
the intrinsic alignment signal and subtract it, or one tries to
remove it altogether.   The former approach is not practical, as,
although there is some agreement as to the general level of the
contamination, the details are not accurately enough known.  The
latter approach is becoming possible, as lensing surveys are now
obtaining estimates of the distance to each galaxy, via
photometric redshifts (spectroscopic redshifts are difficult to
obtain, because one needs a rather deep sample, with median
redshift at least 0.6 or so, and large numbers, to reduce shot
noise due to the random orientations of ellipticities).   With
photometric redshifts, one can remove physically close galaxies
from the pair statistics (such as the shear correlation
function)\cite{HH,KS}. Thus one removes a systematic error in
favour of a slightly increased statistical error.  The analysis in
\cite{Heymans} is the only study which has explicitly removed
close pairs.  

Another effect which is more subtle is the correlation between the orientation or a foreground galaxy and the orientation of the  
lensed image of a background galaxy.  The latter is affected gravitationally by the tidal field in the vicinity of the former (as well as the tidal fields all the way along the line-of-sight), and if the orientation of the foreground galaxy is affected by the local tidal field, as it surely must at some level, then there can be a contamination of the cosmological lensing signal by what is sometimes referred to as the `GI correlation'.  This was first pointed out in \cite{HS04}, and it seems to be likely to be a significant effect \cite{Heymans06}.  It is less easy to deal with than the II correlation, but modelling and nulling methods exist, at the price of diminished signal-to-noise \cite{BK,JS}.

\begin{figure}
\sidecaption
\includegraphics[height=6cm]{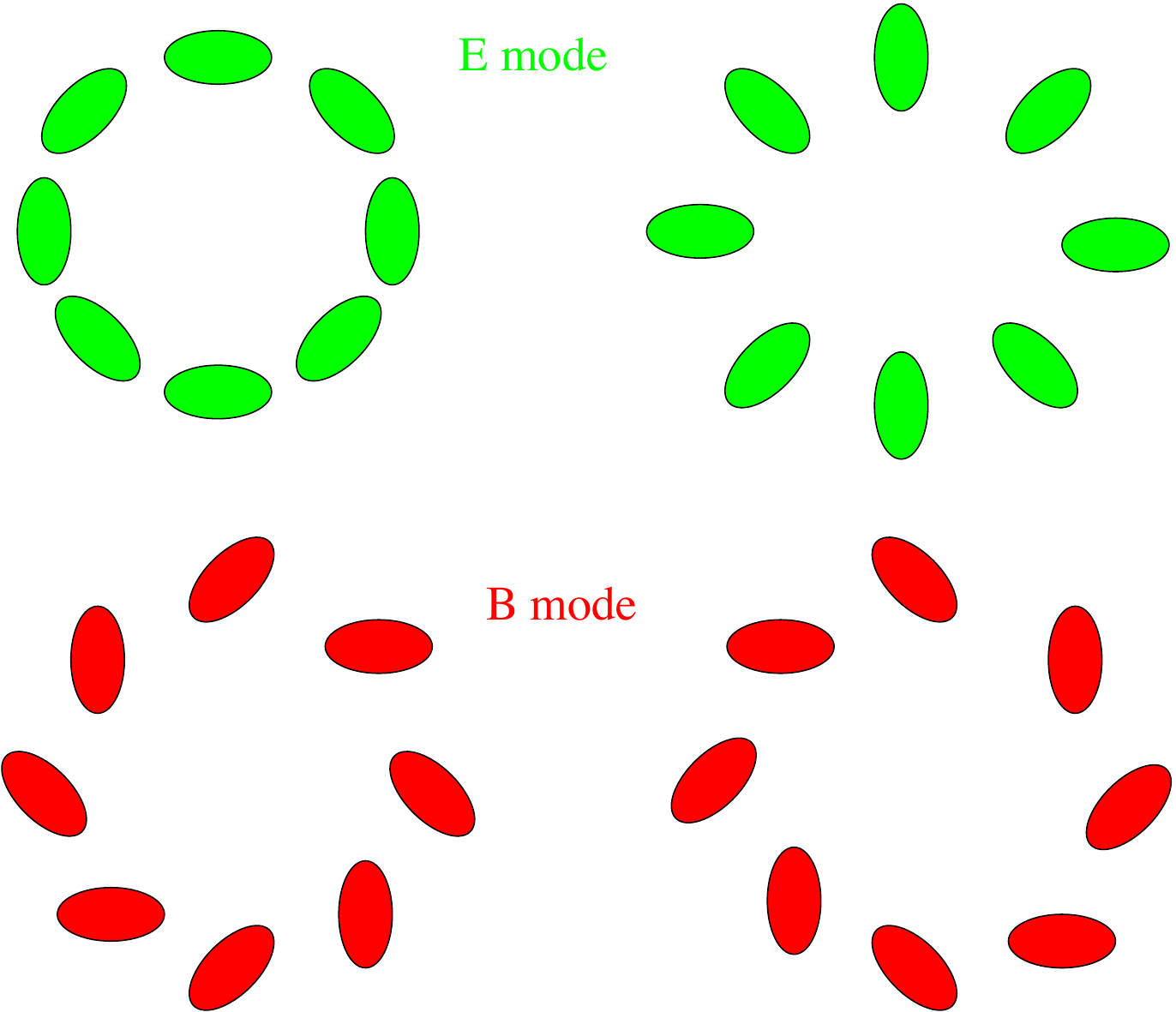}
%
%
\caption{Example patterns from E-mode and B-mode fields
(from \cite{VWM}). Weak lensing only produces E-modes at any
significant level, so the presence of B-modes can indicate
systematic errors.}
\label{EB}       
\end{figure}

\subsection{E/B decomposition}
\label{E/B decomposition}

Weak gravitational lensing does not produce the full range of locally
linear distortions possible.  These are characterised by translation,
rotation, dilation and shear, with six free parameters.  Translation
is not readily observable, but weak lensing is specified by three
parameters rather than the four remaining degrees of freedom permitted
by local affine transformations. This restriction is manifested in a
number of ways: for example, the transformation of angles involves a
$2 \times 2$ matrix which is symmetric, so is not completely general, see
equation (\ref{defA}). Alternatively, a general spin-weight 2 field
can be written in terms of second derivatives of a {\em complex}
potential, whereas the lensing potential is real. There are many other consistency relations which have to
hold if lensing is responsible for the observed shear field. In
practice the observed ellipticity field may not satisfy the expected
relations, if it is contaminated by distortions not associated with
weak lensing.  The most obvious of these is optical distortions of the
telescope system, but could also involve physical effects such as
intrinsic alignment of galaxy ellipticities, which we will consider later.

A convenient way to characterise the distortions is via E/B
decomposition, where the shear field is described in terms of an
`E-mode', which is allowed by weak lensing, and a `B-mode', which is
not.  These terms are borrowed from similar decompositions in
polarisation fields.  In fact weak lensing can generate B-modes, but
they are expected to be very small \cite{Schneider02a}, so
the existence of a significant B-mode in the observed shear pattern
is indicative of some non-lensing contamination.  The easiest way to
introduce a B-mode mathematically is to make the lensing potential
complex:
\begin{equation}
\phi=\phi_E+i\phi_B.
\end{equation}
There are various ways to determine whether a B-mode is present. A
neat way is to generalise a common statistic called the aperture mass to a complex
$M=M_{ap}+iM_\perp$, where the real part picks up the E modes, and
the imaginary part the B modes.  Alternatively, the $\xi_\pm$ can be
used \cite{Crittenden02,Schneider02b}:
\begin{equation}
P_{\kappa\pm}(\ell) = \pi \int_0^\infty d\theta\,
\theta\,[J_0(\ell\theta)\xi_+(\theta)\pm
J_4(\ell\theta)\xi_-(\theta)]
\end{equation}
where the $\pm$ power spectra refer to E and B mode powers. In
principle this requires the correlation functions to be known over
all scales from $0$ to $\infty$. Variants of this (see e.g.
\cite{Crittenden02}) allow the E/B-mode correlation functions to
be written in terms of integrals of $\xi_\pm$ over a finite range:
\begin{eqnarray}
\xi_E(\theta)&=&{1\over
2}\left[\xi_-(\theta)+\xi'_+(\theta)\right]\\\nonumber
\xi_B(\theta)&=&-\frac{1}{2}\left[\xi_-(\theta)-\xi'_+(\theta)\right],
\end{eqnarray}
where
\begin{equation}
\xi'_+(\theta)=\xi_+(\theta)+4\int_0^\theta
\frac{d\theta'}{\theta'}\xi_+(\theta') -12\theta^2\int_0^\theta
\frac{d\theta'}{{\theta'}^3}\xi_+(\theta').
\end{equation}
This avoids the need to know the correlation functions on large
scales, but needs the observed correlation functions to be
extrapolated to small scales; this was one of the approaches taken
in the analysis of the CFHTLS data \cite{Hoekstra06}.
Difficulties with estimating the correlation functions on small
scales have led others to prefer to extrapolate to large scales,
such as in the analysis of the GEMS \cite{Heymans05} and William
Herschel data \cite{Massey05}.  Note that without full sky
coverage, the decomposition into E and B modes is ambiguous,
although for scales much smaller than the survey it is not an issue.

\subsection{Results}

\begin{figure}
\sidecaption
\includegraphics[height=5cm]{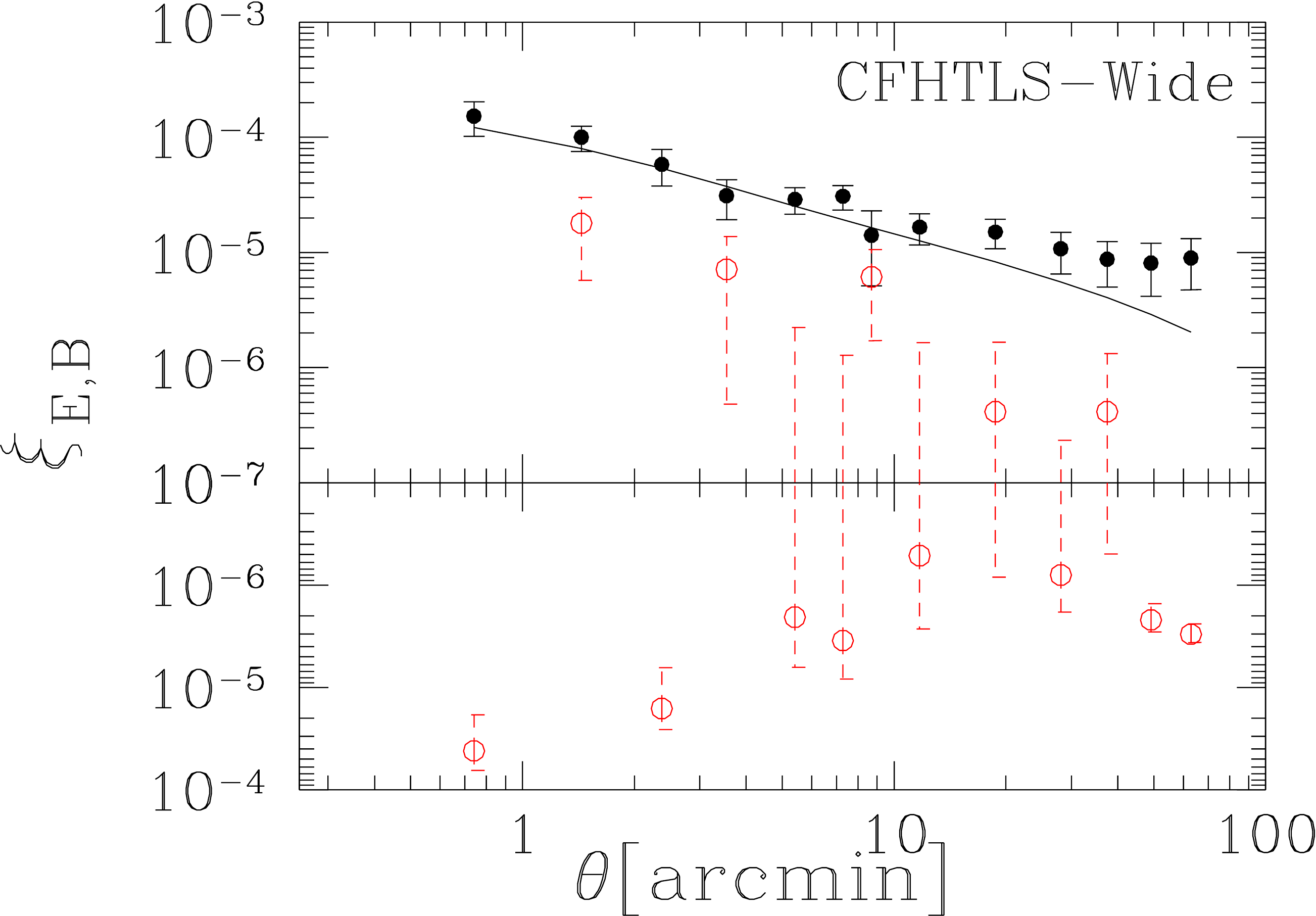}
\caption{E- and B-modes from an early analysis of CFHTLS data \cite{Benjamin}.   Top points are the E-modes and bottom the
B-modes.}
\label{CFHTLS}       
\end{figure}

\begin{figure}
\sidecaption
\includegraphics[height=6cm]{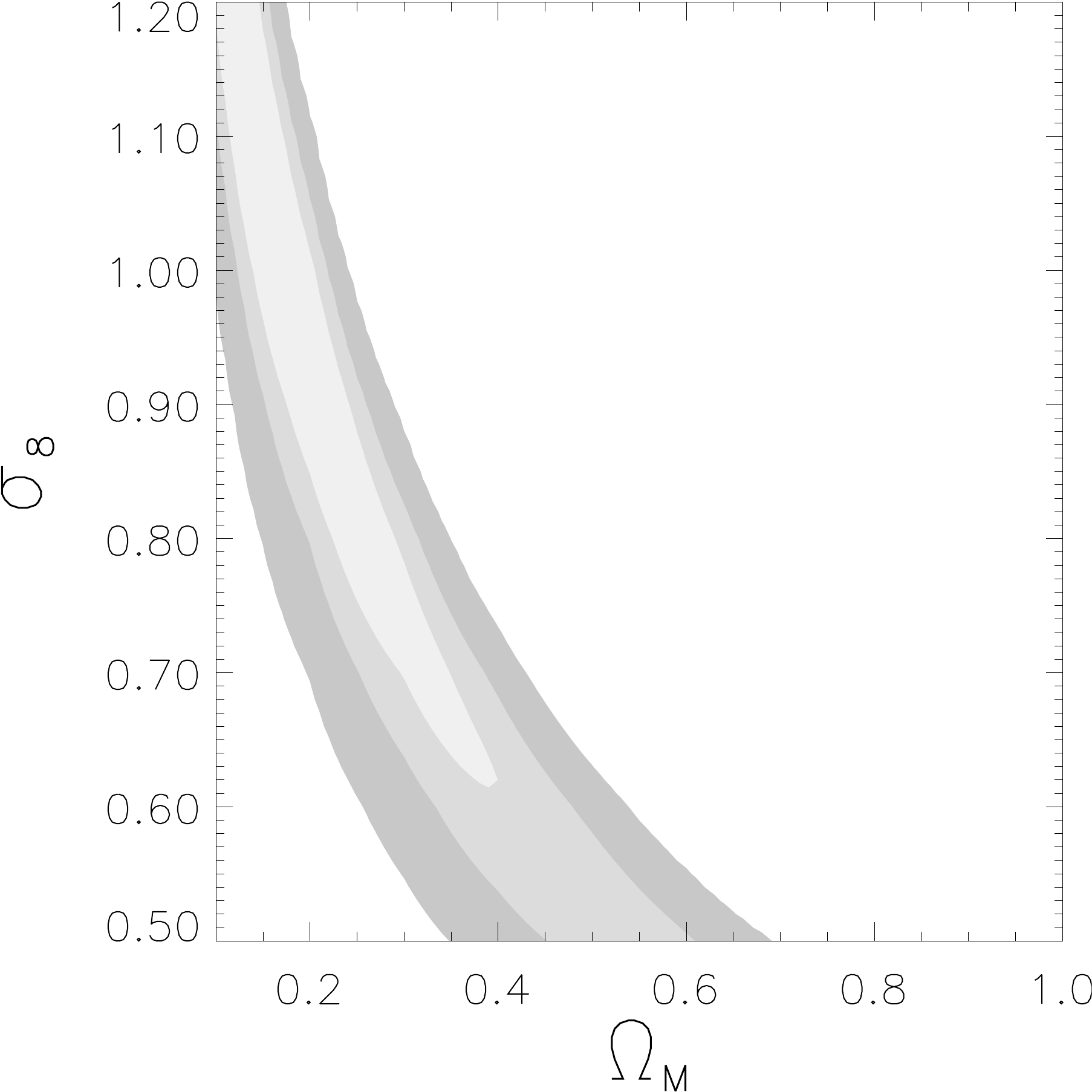}
\caption{Cosmological parameters from an early analysis of CFHTLS data \cite{Benjamin}}
\label{COSMO}       
\end{figure}

The first results from cosmic shear were published in
2000 \cite{Bacon,vWetal,Kaiser,Wittman}, so as an observational
science, cosmological weak lensing is very young. To date, the
surveys have been able to show clear detections of the effect, and
reasonably accurate determination of some cosmological parameters,
usually the amplitude of the dark matter perturbations (measured
by the rms fractional fluctuations in an 8$h^{-1}Mpc$ sphere and
denoted $\sigma_8$), and the matter density parameter $\Omega_m$.
Current surveys cannot lift a near-degeneracy between these two,
and usually a combination (typically $\sigma_8 \Omega_m^{0.5}$) is
quoted.  This makes sense - it is difficult, but certainly not
impossible, to distinguish between a highly-clumped low-density
universe, and a modestly-clumped high-density universe.  There is
no question that the surveys do not yet have the size, nor the
careful control of systematics required to compete with the
microwave background and other techniques used for cosmological
parameter estimation. However, this situation is changing fast,
particularly with the CFHT Legacy Survey, which is now complete, and Pan-STARRS 1, which is underway.  Future surveys such as proposed by
Euclid, JDEM and LSST should be much more powerful.
Early results from CFHTLS are shown in Fig. \ref{CFHTLS} and \ref{COSMO}. 

\section{Lensing in 3D}

Knowing the redshift distribution of sources is vital to interpret 2D lensing statistics, and to help with this, recent lensing surveys have
obtained multi-colour data to estimate distances using photo-zs.  With photo-zs for sources, it makes sense to use the information individually, and this opens up the possibility of 3D lensing.

\subsection{3D potential and mass density reconstruction}

As we have already seen, it is possible to reconstruct the surface
density of a lens system by analysing the shear pattern of galaxies
in the background.  An interesting question is then whether the
reconstruction can be done in three dimensions, when distance
information is available for the sources.     It is probably
self-evident that mass distributions can be {\em constrained} by the
shear pattern, but the more interesting possibility is that one may
be able to {\em determine} the 3D mass density in an essentially
non-parametric way from the shear data.

The idea \cite{Taylor} is that the shear pattern is derivable from
the lensing potential $\phi({\bf r})$, which is dependent on the
gravitational potential $\Phi({\bf r})$ through the integral
equation
\begin{equation}
\phi({\bf r}) = {2\over c^2}\int_0^r \,dr' \left({1\over r'}-{1\over
r}\right) \Phi({\bf r'})\label{phi}
\end{equation}
where the integral is understood to be along a radial path (the Born
approximation), and a flat Universe is assumed in equation
(\ref{phi}).  The gravitational potential is related to the density
field via Poisson's equation (\ref{Poisson}).  There are two problems
to solve here; one is to construct $\phi$ from the lensing data, the
second is to invert equation (\ref{phi}).  The second problem is
straightforward: the solution is
\begin{equation}
\Phi({\bf r}) = {c^2\over 2}{\partial\over \partial
r}\left[r^2{\partial\over \partial r} \phi({\bf r})\right].
\end{equation}
From this and Poisson's equation $\nabla^2\Phi = (3/2)H_0^2\Omega_m
\delta /a(t)$, we can reconstruct the mass overdensity field
\begin{equation}
\delta({\bf r}) = {a(t)c^2\over 3 H_0^2 \Omega_m}
\nabla^2\left\{{\partial\over
\partial r}\left[r^2{\partial\over \partial r} \phi({\bf r})\right]\right\}.
\end{equation}
The construction of $\phi$ is more tricky, as it is not directly
observable, but must be estimated from the shear field. This
reconstruction of the lensing potential suffers from a similar
ambiguity to the mass-sheet degeneracy for simple lenses. To see
how, we first note that the complex shear field $\gamma$ is the
second derivative of the lensing potential:
\begin{equation}
\gamma({{\bf r}}) = \left[{1\over 2}\left({\partial^2\over \partial \theta_x^2}-
{\partial^2\over \partial \theta_y^2}\right) + i {\partial^2\over
\partial \theta_x \partial \theta_y}\right]\phi({\bf r}).
\end{equation}
As a consequence, since the lensing potential is real, its estimate
is ambiguous up to the addition of any field $f({\bf r})$ for which
\begin{equation}
{\partial^2 f({\bf r})\over \partial \theta_x^2}- {\partial^2 f({\bf r})\over
\partial \theta_y^2} = {\partial^2 f({\bf r})\over
\partial \theta_x \partial \theta_y} = 0.
\end{equation}
Since $\phi$ must be real, the general solution to this is
\begin{equation}
f({\bf r}) = F(r) + G(r)\theta_x + H(r)\theta_y + P(r)(\theta_x^2 + \theta_y^2)
\end{equation}
where $F$, $G$, $H$ and $P$ are arbitrary functions of $r\equiv
|{\bf r}|$.  Assuming these functions vary smoothly with $r$, only the
last of these survives at a significant level to the mass density,
and corresponds to a sheet of overdensity
\begin{equation}
\delta = {4 a(t) c^2\over 3H_0^2 \Omega_m r^2} {\partial\over
\partial r} \left[r^2
{\partial\over
\partial r}P(r)\right].
\end{equation}
There are a couple of ways to deal with this problem.  For a
reasonably large survey, one can assume that the potential and its
derivatives are zero on average, at each $r$, or that the
overdensity has average value zero.  For further details, see \cite{BT}.  Note that the relationship between the
overdensity field and the lensing potential is a linear one, so if
one chooses a discrete binning of the quantities, one can use
standard linear algebra methods to attempt an inversion, subject to
some constraints such as minimising the expected reconstruction
errors.  With prior knowledge of the signal properties, this is the
Wiener filter.  See \cite{HuK} for further details of this
approach.

\begin{figure}
\sidecaption
\includegraphics[height=6cm]{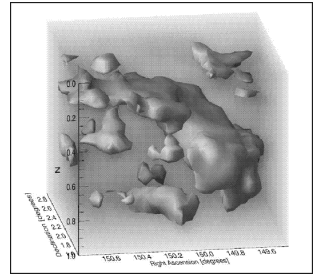}
\caption{3D reconstruction of matter density from the COSMOS ACS data \cite{Massey07}.}
\label{COSMOS3D}
\end{figure}

This method was first applied to COMBO-17 data \cite{Taylor04}, and recently to COSMOS HST data \cite{Massey07} - see Fig.\ref{COSMOS3D}.

\subsection{\label{tomography}Tomography}

In the case where one has distance information for individual
sources, it makes sense to employ the information for statistical
studies.  A natural course of action is to divide the survey into
slices at different distances, and perform a study of the shear
pattern on each slice.  In order to use the information effectively,
it is necessary to look at cross-correlations of the shear fields in
the slices, as well as correlations within each slice \cite{Hu99}.
This procedure is usually referred to as tomography, although the
term does not seem entirely appropriate.

We start by considering the average shear in a shell, which is
characterised by a probability distribution for the source redshifts
$z=z(r)$,  $p(z)$.  The shear field is the second edth derivative of
the lensing potential \cite{CHK}
\begin{equation}
\gamma({\bf r}) = {1\over 2}\edth\edth \phi({\bf r}) \simeq {1\over
2}(\partial_x+i\partial_y)^2\phi({\bf r})
\end{equation}
where the derivatives are in the angular direction, and the last
equality holds in the flat-sky limit. If we average the shear in a
shell, giving equal weight to each galaxy, then the average shear
can be written in terms of an effective lensing potential
\begin{equation}
\phi_{\rm eff}({\bf \theta}) = \int_0^\infty dz\, p(z) \phi({\bf r})
\end{equation}
where the integral is at fixed ${\bf\theta}$, and $p(z)$ is zero
outside the slice (we ignore errors in distance estimates such as
photometric redshifts; these could be incorporated with a suitable
modification to $p(z)$).  In terms of the gravitational potential,
the effective lensing potential is
\begin{equation}
\phi_{\rm eff}({\bf \theta}) = {2\over c^2}\int_0^\infty dr
\,\Phi({\bf r}) g(r)
\end{equation}
where reversal of the order of integration gives the lensing
efficiency to be
\begin{equation}
g(r) = \int_{z(r)}^\infty dz'\, p(z') \left({1\over r}-{1\over
r'}\right),
\label{gr}
\end{equation}
$z'=z'(r')$ and we assume flat space.  If we perform a
spherical harmonic transform of the effective potentials for slices
$i$ and $j$, then the cross power spectrum can be related to the
power spectrum of the gravitational potential $P_\Phi(k)$ via a
version of Limber's equation:
\begin{equation}
\langle \phi^{(i)}_{\ell m}\phi^{*(j)}_{\ell' m'}\rangle =
C^{\phi\phi}_{\ell,ij}\, \delta_{\ell'\ell}\delta_{m'm}
\end{equation}
where
\begin{equation}
C^{\phi\phi}_{\ell,ij}=\left({2\over c^2}\right)^2 \int_0^\infty
dr\,\, {g^{(i)}(r)g^{(j)}(r)\over r^2}\, P_\Phi(\ell/r;r)
\label{CovTom}
\end{equation}
is the cross power spectrum of the lensing potentials. The last
argument in $P_\Phi$ allows for evolution of the power spectrum with
time, or equivalently distance. The power spectra of the convergence
and shear are related to $C^{\phi\phi}_{\ell,ij}$ by \cite{Hu00}
\begin{eqnarray}
C^{\kappa\kappa}_{\ell,ij}&=&{\ell^2(\ell+1)^2\over
4}\,C^{\phi\phi}_{\ell,ij}\\\nonumber
C^{\gamma\gamma}_{\ell,ij}&=&{1\over 4}{(\ell+2)!\over
(\ell-2)!}\,C^{\phi\phi}_{\ell,ij}.
\end{eqnarray}
The sensitivity of the cross power spectra to cosmological
parameters is through various effects, as in 2D lensing: the shape
of the linear gravitational potential power spectrum is dependent on
some parameters, as is its nonlinear evolution; in addition the
$z(r)$ relation probes cosmology, via 
\begin{equation}
r(z) = c\int_0^\infty \frac{dz'}{H(z}.
\end{equation}
The reader is referred to
standard cosmological texts for more details of the dependence of
the distance-redshift relation on cosmological parameters.

\begin{figure}
\sidecaption
\includegraphics[width=6cm, angle=0]{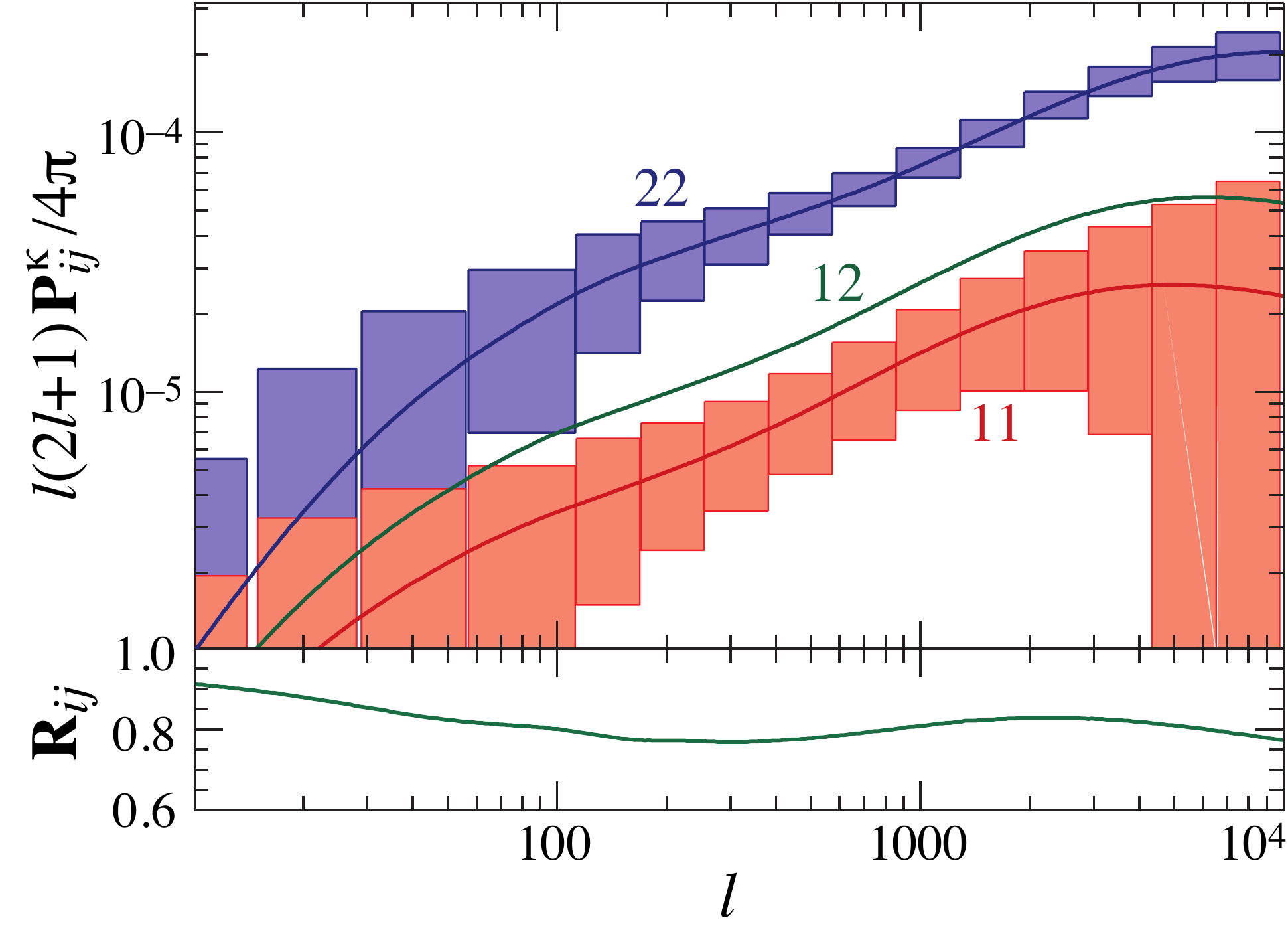}
\caption{The power spectra of two slices, their cross power
spectrum, and their correlation coefficient.  From \cite{Hu99}}
\label{Hutomography}
\end{figure}
Hu \cite{Hu99} illustrates the power and limitation of tomography, with
two shells (Fig. \ref{Hutomography}).  As expected, the deeper shell
(2) has a larger lensing power spectrum than the nearby shell (1),
but it is no surprise to find that the power spectra from shells are
correlated, since the light from both passes through some common
material.  Thus one does gain from tomography, but, depending on
what one wants to measure, the gains may or may not be very much.
For example, tomography adds rather little to the accuracy of the
amplitude of the power spectrum, but far more to studies of dark
energy properties.  One also needs to worry about systematic
effects, as leakage of galaxies from one shell to another, through
noisy or biased photometric redshifts, can degrade the accuracy of
parameter estimation \cite{Huterer06, Ma06, AR08,Kitching08,Kitching09}.

\subsection{\label{shearRatio}The Shear Ratio test}

The shear contributed by the general large-scale structure is
typically about 1\%, but the shear behind a cluster of galaxies can
far exceed this.  As always, the shear of a background source is
dependent on its redshift, and on cosmology, but also on the mass
distribution in the cluster.  This can be difficult to model, so it
is attractive to consider methods which are decoupled from the
details of the mass distribution of the cluster.  Various methods
have been proposed (e.g. \cite{JT,BernsteinJain04}). The method currently receiving the most
attention is simply to take ratios of average tangential shear in
different redshift slices for sources behind the cluster.

The amplitude of the induced tangential shear is dependent on the
source redshift $z$, and on cosmology via the angular diameter
distance-redshift relation $S_k[r(z)]$ by \cite{Taylor07}
\begin{equation}
\gamma_t (z) = \gamma_t(z=\infty){S_k[r(z)-r(z_l)]\over S_k[r(z)]},
\end{equation}
where $\gamma_{t,\infty}$ is the shear which a galaxy at infinite
distance would experience, and which characterises the strength of
the distortions induced by the cluster, at redshift $z_l$.
Evidently, we can neatly eliminate the cluster details by taking
ratios of tangential shears, for pairs of shells in source redshift:
\begin{equation}
R_{ij} \equiv {\gamma_{t,i}\over \gamma_{t,j}}
={S_k[r(z_j)]\,S_k[r(z_i)-r(z_l)]\over
S_k[r(z_i)]\,S_k[r(z_j)-r(z_l)]}.
\end{equation}
In reality, the light from the more distant shell passes through an
extra pathlength of clumpy matter, so suffers an additional source
of shear.  This can be treated as a noise term \cite{Taylor07}.
This approach is attractive in that it probes cosmology through the
distance-redshift relation alone, being (at least to good
approximation) independent of the growth rate of the fluctuations.
Its dependence on cosmological parameters is therefore rather
simpler, as many parameters (such as the amplitude of matter
fluctuations) do not affect the ratio except through minor
side-effects.  More significantly, it can be used in conjunction
with lensing methods which probe both the distance-redshift relation
and the growth-rate of structure. Such a dual approach can in
principle distinguish between quintessence-type dark energy models
and modifications of Einstein gravity.  This possibility arises
because the effect on global properties (e.g. $z(r)$) is different
from the effect on perturbed quantities (e.g. the growth rate of the
power spectrum) in the two cases.  The method has a signal-to-noise
which is limited by the finite number of clusters which are massive
enough to have measurable tangential shear.  In an all-sky survey,
the bulk of the signal would come from the $10^5-10^6$ clusters
above a mass limit of $10^{14}M_\odot$.

\subsection{\label{3Danalysis}Full 3D analysis of the shear field}

An alternative approach to take is to recognise that, with
photometric redshift estimates for individual sources, the data one
is working with is a very noisy 3D shear field, which is sampled at
a number of discrete locations, and for whom the locations are
somewhat imprecisely known.  It makes some sense, therefore, to deal
with the data one has, and to compare the statistics of the discrete
3D field with theoretical predictions.  This was the approach of
\cite{Heavens03, CHK, Heavens06, Kitching07}.  It
should yield smaller statistical errors than tomography, as it
avoids the binning process which loses information.

In common with many other methods, one has to make a decision
whether to analyse the data in configuration space or in the
spectral domain.  The former, usually studied via correlation
functions, is advantageous for complex survey geometries, where the
convolution with a complex window function implicit in spectral
methods is avoided.  However, the more readily computed correlation
properties of a spectral analysis are a definite advantage for
Bayesian parameter estimation, and we follow that approach here.

The natural expansion of a 3D scalar field $(r,\theta,\phi)$ which
is derived from a potential is in terms of products of spherical
harmonics and spherical Bessel functions, $j_\ell(kr)
Y_\ell^m(\vth)$.  Such products, characterised by 3 spectral
parameters $(k,\ell,m)$, are eigenfunctions of the Laplace operator,
thus making it very easy to relate the expansion coefficients of the
density field to that of the potential (essentially via $-k^2$ from
the $\nabla^2$ operator). Similarly, the 3D expansion of the lensing
potential,
\begin{equation}
\phi_{\ell m}(k) \equiv \sqrt{2\over \pi} \int d^3{\bf r}\,
\phi({\bf r}) k j_\ell(kr) Y_\ell^m(\vth),
\end{equation}
where the prefactor and the factor of $k$ are introduced for
convenience. The expansion of the complex shear field is most
naturally made in terms of spin-weight 2 spherical harmonics
$\phantom{.}_2Y_\ell^m$ and spherical Bessel functions, since
$\gamma={1\over 2}\edth\edth\phi$, and $\edth\edth Y_\ell^m \propto
\,\phantom{.}_2Y_\ell^m$:
\begin{equation}
\gamma({\bf r})=\sqrt{2\pi}\sum_{\ell m}\int dk\, \gamma_{\ell m}
\,k\,j_\ell(kr)\, \phantom{.}_2Y_\ell^m(\vth).
\end{equation}
The choice of the expansion becomes clear when we see that the
coefficients of the shear field are related very simply to those of
the lensing potential:
\begin{equation}
\gamma_{\ell m}(k) = {1\over 2}\sqrt{(\ell+2)!\over (\ell-2)!} \,\,
\phi_{\ell m}(k).
\end{equation}
The relation of the $\phi_{\ell m}(k)$ coefficients to the expansion
of the density field is readily computed, but more complicated as
the lensing potential is a weighted integral of the gravitational
potential.  The details will not be given here, but relevant effects
such as photometric redshift errors, nonlinear evolution of the
power spectrum, and the discreteness of the sampling are easily
included. The reader is referred to the original papers for details.

In this way the correlation properties of the $\gamma_{\ell m}(k)$
coefficients can be related to an integral over the power spectrum,
involving the $z(r)$ relation, so cosmological parameters can be
estimated via standard Bayesian methods from the coefficients.
Clearly, this method probes the dark energy effect on both the
growth rate and the $z(r)$ relation.

\subsection{\label{forecasts}Dark Energy with 3D lensing methods}

In this section we summarise some of the forecasts for cosmological
parameter estimation from 3D weak lensing.  We concentrate on the
statistical errors which should be achievable with the shear ratio
test and with the 3D power spectrum techniques.  Tomography should
be similar to the latter.  We show results from 3D weak lensing
alone, as well as in combination with other experiments.  These
include CMB, supernova and baryon oscillation studies.  The methods
generally differ in the parameters which they constrain well, but
also in terms of the degeneracies inherent in the techniques.  Using
more than one technique can be very effective at lifting the
degeneracies, and very accurate determinations of cosmological
parameters, in particular dark energy properties, may be achievable
with 3D cosmic shear surveys covering thousands of square degrees of
sky to median source redshifts of order unity.

\begin{figure}\label{4Exp}
\sidecaption
\includegraphics[width=6cm, angle=0]{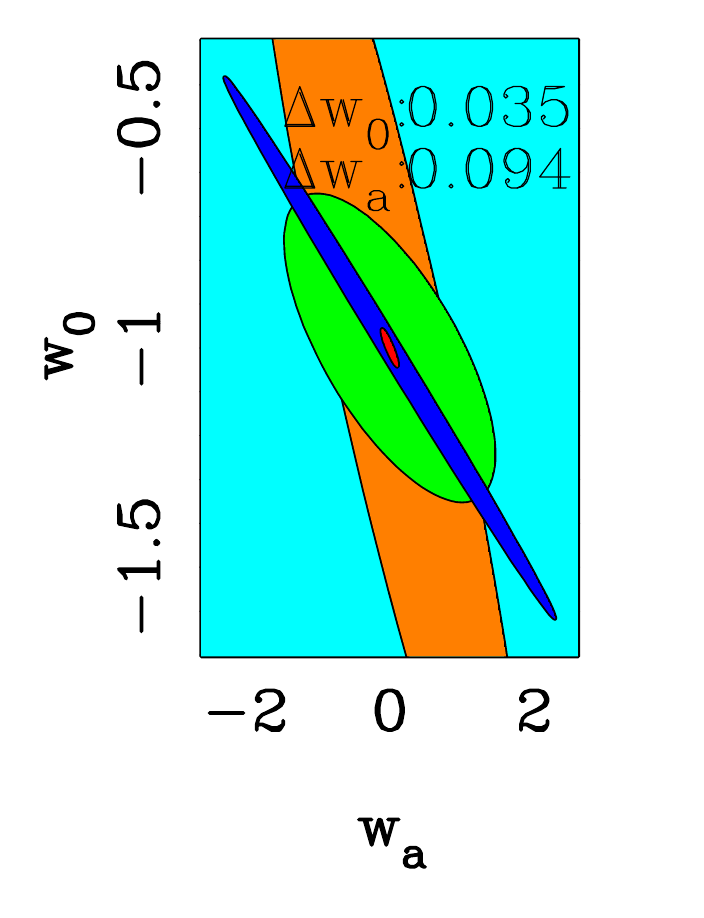}
\caption{The accuracy expected from the combination of experiments
dedicated to studying Dark Energy properties.   Marginal 1$\sigma$, 2-parameter regions are shown
for the experiments individually and in combination.  The supernova
study fills the plot, the thin diagonal band is Planck, the
near-vertical band is BAO, and the ellipse is the 3D lensing power
spectrum method.  The small ellipse is from combined. From \cite{Heavens06}.}
\end{figure}

\begin{figure}
\sidecaption
\includegraphics[width=6cm, angle=0]{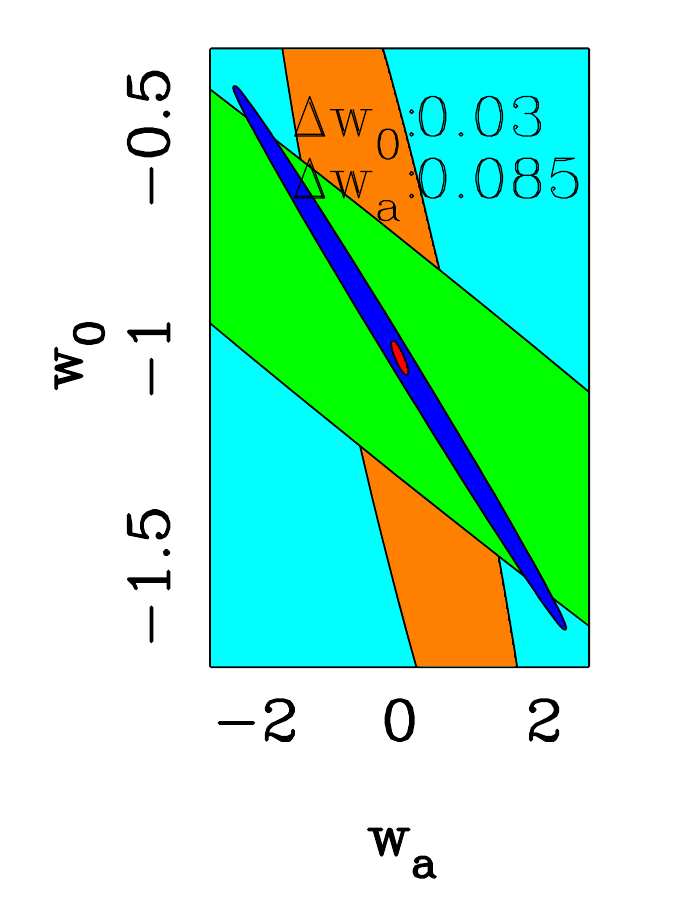}
\caption{As in Fig. \ref{4Exp}, but with the shear ratio test as the
lensing experiment.  Supernovae fill the plot, Planck is the thin
diagonal band, BAO the near-vertical band, and the shear ratio is
the remaining 45 degree band.  The combination of all experiments is
in the centre.   From \cite{Taylor07}.}
\label{all}
\end{figure}

Fig. \ref{4Exp}  and \ref{all} show the accuracy which might be achieved with a number
of surveys designed to measure cosmological parameters.  We
concentrate here on the capabilities of each method, and the methods
in combination, to constrain the dark energy equation of state, and
its evolution, parametrised by \cite{CP01}
\begin{equation}
w(a) = {p\over \rho c^2} = w_0 + w_a(1-a)
\end{equation}
where the behaviour as a function of scale factor $a$ is, in the
absence of a compelling theory, assumed to have this simple form.
$w=-1$ would arise if the dark energy behaviour was actually a
cosmological constant.

The assumed experiments are: a 5-band 3D weak lensing survey,
analysed either with the shear ratio test, or with the spectral
method, covering 10,000 square degrees to a median redshift of 0.7,
similar to the capabilities of a groundbased 4m-class survey with a
several square degree field; the Planck CMB experiment (14-month
mission); a spectroscopic survey to measure baryon oscillations
(BAO) in the galaxy matter power spectrum, assuming constant bias,
and covering 2000 square degrees to a median depth of unity, and a
smaller $z=3$ survey of 300 square degrees, similar to WFMOS
capabilities on Subaru; a survey of 2000 Type Ia supernovae to
$z=1.5$, similar to SNAP's design capabilities.

We see that the experiments in combination are much more powerful
than individually, as some of the degeneracies are lifted.  Note
that the combined experiments appear to have rather smaller error
bars than is suggested by the single-experiment constraints.  This
is because the combined ellipse is the projection of the product of
several multi-dimensional likelihood surfaces, which intersect in a
small volume.  (The projection of the intersection of two surfaces
is not the same as the intersection of the projection of two
surfaces). The figures show that errors of a few percent on $w_0$
are potentially achievable, or, with this parametrisation, an error
of $w$ at a `pivot' redshift of $z\simeq 0.4$ of under 0.02.  This
error is essentially the minor axis of the error ellipses.  These graphs include statistical errors only;  systematic errors, from, for example bias in the photo-z distribution, are expected to degrade errors by $\sim \sqrt{2}$ \cite{Kitching08}.

\section{Dark Gravity}

In addition to the possibility that Dark Energy or the cosmological constant drives acceleration, there is an even more radical solution.  As a cosmological constant, Einstein's term represents a modification of the gravity law, so it is interesting to consider whether the acceleration may be telling us about a failure of GR.  Although no compelling theory currently exists, suggestions include modifications arising from extra dimensions, as might be expected from string-theory braneworld models.  Interestingly, there are potentially measurable effects of such exotic gravity models which weak lensing can probe, and finding evidence for extra dimensions would of course signal a radical departure from our conventional view of the Universe.

Weak lensing is useful as it probes not just the distance-redshift relation, but also the growth rate of perturbations (see, for example, equations \ref{gr} and \ref{CovTom}).  This is important, because in principle measures which use only the $r(z)$ relation suffer from a degeneracy between a different gravity law and the equation of state of the contents of the Universe.  To see this, first note that a modified gravity law will lead to some sort of Hubble relation $H(a)$.  If we combine the Friedmann equation
\begin{equation}
H^2(a) + \frac{k}{a^2} = \frac{8\pi G \rho}{3}
\end{equation}
and
\begin{equation}
\frac{d}{da}\left(\rho a^3\right) = -p a^2 = - w(a)\rho a^2
\end{equation}
we find that
\begin{equation}
w(a) = -\frac{1}{3}\frac{d}{d\ln a}\left[\frac{1}{\Omega_m(a)}-1\right].
\end{equation}
So we see that any modified gravity law can be mimicked, as far as the distance-redshift relation is concerned, by GR with an appropriate equation of state.

\begin{figure}
\sidecaption
\includegraphics[width=2.787in]{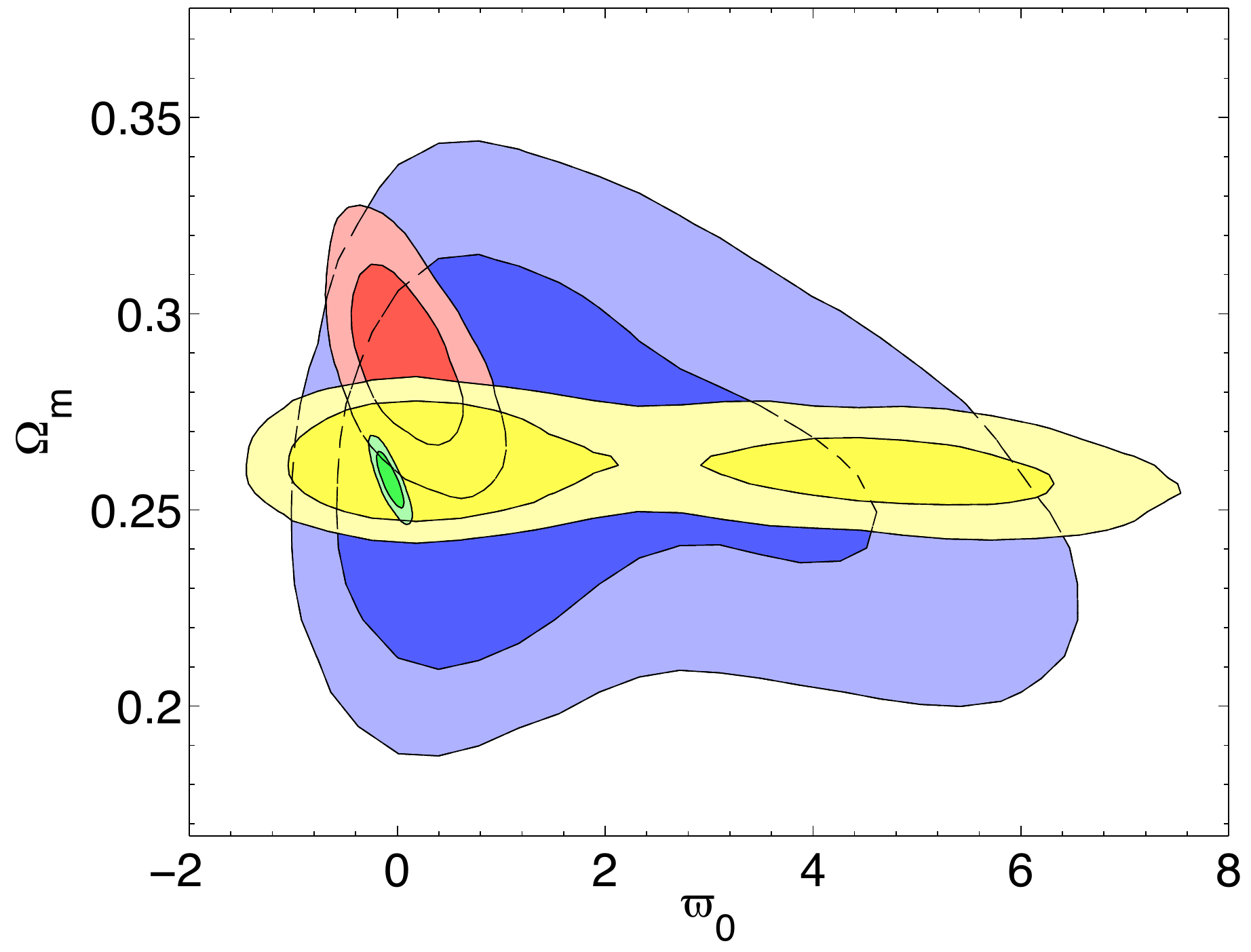}
\caption{The projected marginal 68\% and 95\% likelihood contours for the slip, $\varpi$, assuming $\varpi=\varpi_0(1+z)^{-3}$, for WMAP 5-year data (blue), adding current weak lensing and ISW data (red).  Yellow is mock Planck CMB data, and green adds weak lensing from a 20,000 square degree survey \cite{Daniel}. 
\label{Daniel}}
\end{figure}

To analyse other gravity laws,  we consider scalar perturbations in the conformal Newtonian gauge (flat for simplicity), $ds^2 = a^2(\eta)\left[(1+2\psi)d\eta^2 - (1-2\phi)d\vec x^2\right]$, where $\psi$ is the potential fluctuation, and $\phi$ the curvature perturbation, $\eta$ being the conformal time.  Information on the gravity law is manifested in these two potentials.  For example in GR and in the absence of anisotropic stresses (a good approximation for epochs when photon and neutrino streaming are unimportant) $\phi=\psi$.   More generally,  the Poisson law may be modified, and the laws for $\psi$ and $\phi$ may differ.  This difference can be characterised \cite{Daniel} by the {\em slip}, $\varpi$.  This may be scale- and time-dependent: $\psi(k,a) = \left[1+\varpi(k,a)\right]\phi(k,a)$,
and the modified Poisson equation may be characterised by  $Q$, an effective change in $G$ \cite{Amendola}:
\begin{equation}
-k^2\phi = 4\pi G a^2 \rho_m \delta_m Q(k,a).
\end{equation}
Different observables are sensitive to $\psi$ and $\phi$ in different ways \cite{JainZhang}.  For example, the Integrated Sachs-Wolfe effect depends on $\dot\psi + \dot\phi$, but the effect is confined to large scales and cosmic variance precludes accurate use for testing modified gravity. Peculiar velocities are sourced by $\psi$.  Lensing is sensitive to $\psi+\phi$, and this is the most promising route for next-generation surveys to probe beyond-Einstein gravity.    The Poisson-like equation for $\psi+\phi$ is
\begin{equation}
-k^2(\psi+\phi) = 2\Sigma\frac{3H_0^2 \Omega_m}{2a} \delta_m
\end{equation}
where $\Sigma \equiv Q(1+\varpi/2)$.   For GR, $\Sigma=1$, $\varpi=0$.  The DGP braneworld model \cite{DGP} has $\Sigma=1$, so mass perturbations deflect light in the same way as GR, but the growth rate of the fluctuations differs.  Thus we have a number of possible observational tests, including probing the expansion history, the growth rate of fluctuations and the mass density-light bending relation.  Future WL surveys can put precise constraints on $\Sigma$ \cite{Amendola}, and on $\varpi$ (see Fig. \ref{Daniel})) \cite{Daniel}.

By probing the growth rate as well as the expansion history, weak lensing can lift a degeneracy which exists in methods which consider the distance-redshift relation alone, since the expansion history in a modified gravity model can always be mimicked by GR and Dark Energy with a suitable $w(a)$.
In general however the growth history of cosmological structures
will be different in the two cases (e.g. \cite{Knox2006,HutererLinder07}, but
see \cite{Kunz}). 

\subsection{Growth rate}
Whilst not the most general, the  growth index $\gamma$ \cite{Linder05} (not to be confused with the shear) is a convenient minimal extension of GR.   
The growth rate of perturbations in the matter density
$\rho_m$, $\delta_m \equiv \delta \rho_m/\rho_m$, is parametrised as a function of scale factor $a(t)$ by
\begin{equation}
\frac{\delta_m}{a} \equiv g(a) =
\exp\left\{\int_0^a\,\frac{da'}{a'}\left[\Omega_m(a')^\gamma-1\right]\right\},
\end{equation}
In the standard GR cosmological model, $\gamma\simeq 0.55$,  whereas in modified gravity 
theories it deviates from this value.  E.g. the flat DGP braneworld model \cite{DGP} has $\gamma\simeq 0.68$
on scales much smaller than those where cosmological 
acceleration is apparent \cite{LinderCahn07}. 

\begin{figure}
\sidecaption
\includegraphics[width=2.787in]{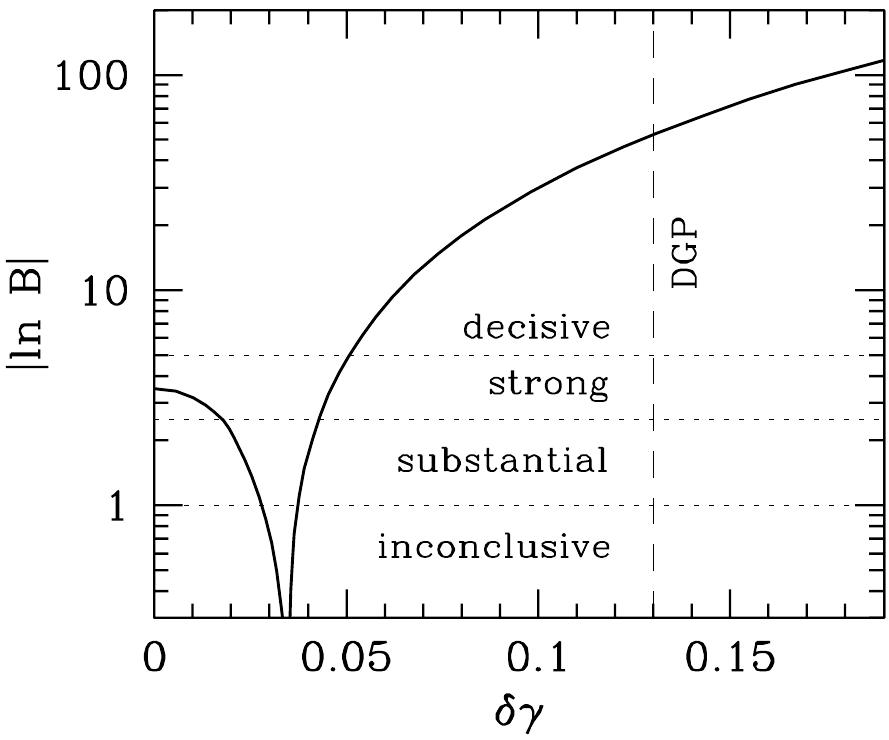}
\caption{Expected Bayesian evidence $B$ vs.
deviation of the growth index from GR, for a future WL survey + \emph{Planck} \cite{HKV}.
If modified gravity is the true model, GR will still be favoured by the data to the left of the cusp.
The Jeffreys scale of evidence \cite{Jeffreys61} is labeled.
\label{dgamma}}
\end{figure}

Measurements of the growth factor  can in principle be used to determine the growth index $\gamma$, and it is interesting to know if it is of any practical use.  In
contrast to parameter estimation, this is an issue of 
model selection - is the gravity model GR, or is there evidence for beyond-Einstein gravity?  This question may 
be answered with the Bayesian evidence, $B$ \cite{Skilling04}, which is the ratio of 
probabilities of two or more models, given some data.
Following  \cite{HKV}, Fig.  \ref{dgamma} shows how the Bayesian evidence for GR changes 
with increasing true deviation of $\gamma$ from its GR value 
for a combination of a future WL survey and \emph{Planck}. 
From the WL data alone, one should be able to distinguish GR decisively 
from the flat DGP model at $\ln B \simeq 11.8$, or, 
in the frequentist view, $5.4\sigma$ \cite{HKV}.
The combination of WL+\emph{Planck}+BAO +SN should be able to distinguish $\delta \gamma = 0.041$ at $3.41$ sigma. This data  combination should be able to decisively distinguish a Dark Energy GR model
from a DGP modified-gravity model with expected
evidence ratio $\ln B \simeq 50$.   An alternative is to parametrise geometry and growth with two separate effective values of $w$, and look for (in)consistency \cite{Ishak,Song,Wang,Zhang}.

One caveat on all of these conclusions is that WL requires knowledge of the nonlinear regime of galaxy clustering, and this is reasonably well-understood for GR, but for other models, further theoretical work is required. This has already started \cite{Schmidt2008}.
The case for a large, space-based 3D weak lensing survey is
strengthened, as it offers the possibility of conclusively
distinguishing Dark Energy from at least some modified gravity
models.

\begin{table}
\begin{center}
\begin{tabular}{|l|c|c|c|}
 \hline
Survey&$\nu$& $|\ln B|$&\\
\hline
DES+\emph{Planck}+BAO+SN& 3.5 &$1.28$&substantial\\
DES+\emph{Planck}& 2.2 &$0.56$&inconclusive\\
DES& 0.7 &$0.54$&inconclusive\\
 \hline
PS1+\emph{Planck}+BAO+SN& 2.9 &$3.78$&strong\\
PS1+\emph{Planck}& 2.6 &$2.04$&substantial\\
PS1& 1.0 &$0.62$&inconclusive\\
 \hline
\emph{WL$_{NG}$}+\emph{Planck}+BAO+SN &10.6&$63.0$&decisive\\
\emph{WL$_{NG}$}+\emph{Planck}& 10.2 &$52.2$&decisive\\
\emph{WL$_{NG}$}& 5.4 &$11.8$&decisive\\
 \hline
\end{tabular}
\caption{The evidence ratio for the three weak lensing experiments
considered with and without \emph{Planck}, supernova and BAO priors.
$WL_{NG}$ is a next-generation space-based imaging survey such as
proposed for \emph{DUNE} or \emph{SNAP}. $z_m$ is the median
redshift, $n_0$ the number of sources per square arcminute, and
$\sigma_z$ is the assumed photometric redshift error. For
completeness, we also list the frequentist significance $\nu\sigma$
with which GR would be expected to be ruled out, if the DGP
braneworld were the correct model. } \label{Results}
\end{center}
\end{table}

\section{The Future}

The main promise of weak lensing in the future will come from
larger surveys with optics designed for excellent image quality.
Currently, the CFHTLS is the state-of-the-art, covering $\sim 170$ square
degrees to a median redshift in excess of one.  In the near future
Pan-STARRS, VST and DES promise very small PSF distortions and
large areal coverage, and in the far future LSST on the ground,
and satellites such as Euclid or JDEM may deliver extremely potent
lensing surveys. In parallel with these developments, the
acquisition of photometric redshifts for the sources has opened up
the exciting possibility of analysing weak lensing surveys in 3D.
Each source represents a noisy estimate of the shear field at a
location in 3D space, and this extra information turns out to be
extremely valuable, increasing substantially the statistical power
of lensing surveys. In particular, it can lift the degeneracy
between $\sigma_8$ and $\Omega_m$, measure directly the growth of
Dark Matter clustering \cite{Bacon04} and, more excitingly still,
it represents a powerful method to measure the equation of state
of Dark Energy \cite{Heavens03,JT,Heavens06} - surely one of the most
important remaining questions in cosmology. In addition,
photometric redshifts allow the possibility of direct 3D Dark
Matter mapping \cite{Taylor,BT,Taylor04}, thus addressing another
of the unsolved problems.  Finally, there is the interesting prospect that we may be able to find evidence for 
extra dimensions in the Universe, from braneworld models, or other cosmological models, from the 
effect they have on the gravity law and the growth rate of perturbations.  Most excitingly, it seems that this may be within reach 
for ambitious lensing surveys planned for the next decade.

\section{Appendix: the propagation of light through a weakly-perturbed
universe}

\subsection{The geodesic equation}

The geodesic equation governs the worldline $x^\lambda$
($\lambda=0,1,2,3$) of a particle, and is readily found in
textbooks on General Relativity (e.g. \cite{dInverno}). It is
\begin{equation}
{d^2x^\lambda\over dp^2} +
\Gamma^\lambda_{\phantom{\lambda}\mu\nu} {dx^\mu\over
dp}{dx^\nu\over dp} =  0 \label{Geodesic0}
\end{equation}
where $p$ is an affine parameter, and $\Aff{\lambda}{\mu}{\nu}$ is
the affine connection, which can be written in terms of the metric
tensor $g_{\mu\nu}$ as
\begin{equation}
\Aff{\lambda}{\mu}{\nu} = {1\over 2} g^{\sigma\lambda}\left\{ {\partial
g_{\mu\nu}\over \partial x^\sigma} +
 {\partial g_{\sigma\nu}\over \partial x^\mu} -
{\partial g_{\mu\sigma}\over \partial x^\nu} \right\}.
\end{equation}
For weak fields, the interval is given by
\begin{equation}
ds^2 = \left(1+{2\Phi\over c^2}\right)c^2 dt^2 -
\left(1-{2\Phi\over c^2}\right)R^2(t)\left[dr^2 +
S_k^2(r)d\beta^2\right]
\end{equation}
where $\Phi$ is the peculiar gravitational potential, $R(t)$ is
the scale factor of the Universe, and $r,\theta,\varphi$ are the
usual comoving spherical coordinates.  The angle $d\beta =
\sqrt{d\theta^2 + \sin^2\theta d\varphi^2}$.  $S_k(r)$ depends on
the geometry of the Universe, being given by
\[
S_k(r)= \left\{
\begin{array}{ll}
  \sinh r, &\mbox{if $k<0$;}\\
  r,       &\mbox{if $k=0$;}\\
  \sin r,  &\mbox{if $k>0$.}
\end{array}\right.
\]
The curvature $k = -1,0,1$ corresponds to open, flat and closed
Universes respectively.  We are interested in the distortion of a
small light bundle, so we can concentrate on a small patch of sky.
If we choose the polar axis of the coordinate system to be along
the centre of the light bundle,  we can define angles $\theta_x
\equiv \theta \cos\varphi$ and $\theta_y\equiv \theta\sin\varphi$.
For convenience we also use the {\em conformal time}, defined by
$d\eta = c dt/R(t)$, in place of the usual time coordinate. With
these definitions the interval is more simply written as
\[
ds^2 = R^2(t)\left\{\left(1+{2\Phi\over c^2}\right)d\eta^2 -
\left(1-{2\Phi\over c^2}\right)\left[dr^2 +
S_k^2(r)(d\theta_x^2+d\theta_y^2)\right]\right\}.
\]
The metric tensor for weakly-perturbed flat
Friedmann-Robertson-Walker metric is then
\[
R^2(t)\left(\begin{array}{cccc}
1+{2\Phi/c^2} & 0 & 0 & 0\\
0 & -\left(1-{2\Phi/c^2}\right) & 0 & 0\\
0 & 0 & -r^2\left(1-{2\Phi/c^2}\right) & 0 \\
0 & 0 & 0 & -r^2\left(1-{2\Phi/c^2}\right) \\
\end{array}\right).
\]
We are interested in how the angles of the ray, $(\theta_x,
\theta_y)$ change as the photon moves along its path, responding
to the varying gravitational potential.  The unperturbed, radial,
path is set by $0 = ds^2 \simeq d\eta^2 - dr^2$, i.e. For a radial
incoming ray,
\[
{dr\over d\eta} = -1.
\]
With $g^{\mu\nu}$ defined as the inverse of $g_{\mu\nu}$ (so
$g_{\mu\nu}g^{\nu\alpha}=\delta_\mu^{\phantom{\mu}\alpha}$), the
affine connections are readily computed.

The parametrized equation for $\eta$ is required only to
zero-order in $\Phi$, and reduces to
\[
{d^2\eta\over dp^2}=-2{\dot R\over R}\dot \eta,
\]
where a dot here denotes $d/dp$.  By choosing the unit of $p$
appropriately, we find
\[
{d\eta\over dp}={1\over R^2}.
\]
We can also relate the radial coordinate to the conformal time,
again to zero-order in $\Phi$: The first-order equations governing
$\theta_x$ and $\theta_y$ are obtained from the geodesic equation,
or by the variational methods (see for example, d'Inverno (1992),
section 7.6 \cite{dInverno})
\[
{\partial L^2\over \partial x^\mu} - {d\over dp}\left({\partial
L^2\over \partial \dot x^\mu}\right) = 0,
\]
where $L^2 \equiv (ds/dp)^2$.  With $x^\mu=\theta_x$,
\[
R^2 {2\over c^2} {\partial \Phi\over \partial \theta_x} \dot
\eta^2 + {2\over c^2}R^2 {\partial \Phi\over \partial \theta_x}
\left(\dot r^2 + r^2 \theta_x^2 + r^2 \theta_y^2\right) -{d\over
dp}\left[-2R^2 r^2 \left(1-{2\Phi\over c^2}\right)\dot
\theta_x\right] = 0.
\]
With the zero-order solutions for $d\eta/dp$ and $dr/d\eta$, to
first order this reduces simply to
\[
{d^2 \theta_x\over d\eta^2} - {2\over r}{d\theta_x\over d\eta} =
-{2\over c^2 r^2}{\partial \Phi\over \partial \theta_x}.
\]
It is convenient to write this as an equation for the comoving
displacement of the ray from a fiducial direction,
\[
x_i \equiv r \theta_i. \qquad i=1,2
\]
and the equation for $\theta_x$ and a similar one for $\theta_y$
simplify to
\[
{d^2 {\bf x} \over d\eta^2} = -{2\over c^2} \nabla\Phi
\]
where $\nabla$ here is a comoving transverse gradient operator
$(\partial_x, \partial_y)$.

We see that the propagation equation for the displacement looks
similar to what one would guess from a Newtonian point-of-view;
the presence of $\eta$ (instead of $t$) in the acceleration term
on the left is a result of the expansion of the Universe and the
choice of comoving coordinates.  The right-hand side looks like
the gradient of the potential, but is larger than the na\i ve
gradient by a factor of two.  This is the same factor of two which
leads to the classic result of General Relativity, famously tested
by Eddington's 1919 solar eclipse observations, that the angle of
light bending by the Sun is double what Newtonian theory
predicted.

\end{document}